\documentclass[12pt]{article}
\usepackage{amsmath}
\usepackage{graphicx}
\usepackage{natbib}
\usepackage{url} 

\newcommand{\blind}{0}

\usepackage[utf8]{inputenc}
\usepackage[T1]{fontenc}
\usepackage{comment}
\usepackage[english]{babel}
\usepackage[top=2.5cm, bottom=2.5cm, left=2.0cm, right=2.0cm]{geometry}
\usepackage{amscd,amsfonts,amsmath,amssymb,amsthm,bbm,bm,latexsym,mathrsfs}
\allowdisplaybreaks
\usepackage{bookmark}
\usepackage{enumitem}
\usepackage{adjustbox}
\usepackage{gensymb}
\usepackage{natbib}
\usepackage{float,rotating}
\usepackage{comment}
\hypersetup{hidelinks}

\usepackage{algorithm}
\usepackage{algorithmic}

\usepackage{caption}
\usepackage{graphicx}
\usepackage{subcaption}
\usepackage{tabu}

\usepackage[usenames]{color}
\usepackage{epsfig,epstopdf,graphics,graphicx,caption,longtable}
\graphicspath{{./figures/}}
\usepackage{tikz}
\usepackage{tkz-berge}
\usetikzlibrary{shapes,arrows,bayesnet}	

\usepackage{color}
\definecolor{France}{RGB}{32,221,60}
\definecolor{Germany}{RGB}{255,77,20}
\definecolor{Italy}{RGB}{50,197,225}
\definecolor{Spain}{RGB}{255,133,20}

\usepackage[toc,page]{appendix}
\usepackage[normalem]{ulem}
\usepackage{cancel}

\makeatother

\usepackage[section]{placeins}
\usepackage{setspace}

\newtheorem{remark}{Remark}[section]

\newtheorem{proposition}{Proposition}[section]

\theoremstyle{remark}

\theoremstyle{plain}

\begin{document}

\def\spacingset#1{\renewcommand{\baselinestretch}%
{#1}\small\normalsize} \spacingset{1}


\if0\blind
{
  \title{\bf A Multiple Random Scan Strategy \\ for Latent Space Models}
  \author{Roberto Casarin\thanks{The authors acknowledge the support from the European Union - Next Generation EU - Project `\textit{GRINS - Growing Resilient, INclusive and Sustainable' project} (PE0000018); the National Recovery and Resilience Plan (NRRP) Spoke 4 and the MUR - PRIN project `\textit{Discrete random structures for Bayesian learning and prediction}' under g.a. n. 2022CLTYP4. The views and opinions expressed are only those of the authors and do not necessarily reflect those of the European Union or the European Commission. Neither the European Union nor the European Commission can be held responsible for them. This research used the HPC multiprocessor cluster system of the the Venice Center for Risk Analytics (VERA) at Ca’ Foscari University of Venice.}\\
   \texttt{r.casarin@unive.it}
   \hspace{.2cm}\\
    VERA Centre, Ca' Foscari University of Venice\\
    and \\
    Antonio Peruzzi \\
    \texttt{antonio.peruzzi@unive.it}\\
    VERA Centre, Ca' Foscari University of Venice}
  \maketitle
 \fi
}
\if1\blind
{
  \bigskip
  \bigskip
  \bigskip
  \begin{center}
    {\LARGE\bf Efficient Approximate Inference for Latent Space Models}
\end{center}
  \medskip
} \fi

\bigskip
\begin{abstract}
Latent Space (LS) network models project the nodes of a network on a $d$-dimensional latent space to achieve dimensionality reduction of the network while preserving its relevant features. Inference is often carried out within a Markov Chain Monte Carlo (MCMC) framework. Nonetheless, it is well-known that the computational time for this set of models increases quadratically with the number of nodes. In this work, we build on the  Random-Scan (RS) approach to propose an MCMC strategy that alleviates the computational burden for LS models while maintaining the benefits of a general-purpose technique. We call this novel strategy Multiple RS (MRS). This strategy is effective in reducing the computational cost by a factor without severe consequences on the MCMC draws. Moreover, we introduce a novel adaptation strategy that consists of a probabilistic update of the set of latent coordinates of each node. Our Adaptive MRS adapts the acceptance rate of the Metropolis step to adjust the probability of updating the latent coordinates. We show via simulation that the Adaptive MRS approach performs better than MRS in terms of mixing. Finally, we apply our algorithm to a multi-layer temporal LS model and show how our adaptive strategy may be beneficial to empirical applications.
\end{abstract}

\noindent%
{\it Keywords:}  Latent Space Models, Random Scan, Adaptive MCMC
\vfill

\newpage
\spacingset{1.75} 
\section{Introduction}
\label{sec:intro}

Latent Space (LS) network models project the nodes of a network on a $d$-dimensional latent space to reduce the network dimensionality and provide an intuitive representation of the similarity between nodes. LS models are nowadays a standard tool in network analysis and have been applied in many fields such as 
biology (\citealp{huang2022latent}),  finance (\citealp{linardi2020dynamic, casarin2024dynamic}), neuroscience (\citealp{durante2017nonparametric, wilson2020hierarchical}), political science (\citealp{barbera2015birds, ParkBA1147, yu2021spatial}) and social science (\citealp{wang2023joint}). The formalization of LS models is due to the seminal work of \citet{hoff2002latent}, (see also \citealp{HoffStatS}). Several extensions of the original model have been proposed. Among others, the original model has been extended to accommodate dynamic (\citealp{friel2016interlocking,sewell2016latent}) and multi-layer settings (\citealp{sosa2022latent}). See \citet{matias2014modeling}, \citet{kim2018review}, and \citet{sosa2021review} for a comprehensive literature review on latent factor models for networks. In its most straightforward representation, the LS model assumes the binary adjacency matrix of the network is parametrized by a set of node-specific latent coordinates. The computational burden of the inference procedure poses substantial limits to the application of these models to large networks. In this paper, we focus on Bayesian inference and the scalability of the numerical methods used for posterior approximation.


Various methods can be used to approximate the likelihood or the posterior distribution. Efficient Variational Inference  (\citealp{salter2013variational}) and Expectation Maximization (\citealp{artico2023dynamic}) have been proposed, although they rely on binary network assumption. In this paper, we focus on the Markov Chain Monte Carlo (MCMC) framework and contribute to improving the scalability of such a method.

It is well known that the MCMC computational time for this set of models increases quadratically with the number of nodes in the network. A first method for reducing the computational cost of LS models has been proposed in \citet{raftery2012fast}. The authors exploit a case-control approximate likelihood for binary networks to reduce the computational burden from  $\mathcal{O}(N^2)$ to $\mathcal{O}(N)$,  where $N$ is the number of nodes. \citet{rastelli2018computationally} reach a computational improvement via a grid approximation of the latent distances, reducing the computational burden to something lower than $\mathcal{O}(N^2)$. \citet{spencer2022faster} proposed a combination of the split Hamiltonian Monte Carlo and Firefly Monte Carlo to achieve computational efficiency in a binary-network setup. In this work, we build on the  Random-Scan (RS) (\citealp{robert1999monte}, \citealp{latuszynski2013adaptive}) approach to propose an MCMC strategy that alleviates the computational burden for LS models while maintaining the benefits of a general-purpose technique, which not only works for binary networks but also for general weighted networks. We call this novel strategy Multiple RS (MRS), consisting of a probabilistic update of multiple latent positions at each iteration. 

The Multiple RS comes at hand especially for those cases in which on-the-fly recentering is applied to latent factors for identification purposes as in the case of LS models (see \citealp[p. 396]{gelman1995bayesian}, \citealp{keefe2018formal} for some examples of on-the-fly recentering). Moreover, we suggest using a novel Adaptive MRS (AMRS) approach based on the acceptance rate of the Metropolis-within-Gibbs. Our adaptive approach can be easily combined with standard Adaptive Metropolis-Hastings (AMH) approaches (\citealp{andrieu2008tutorial, roberts2009examples, latuszynski2013adaptive}). We also provide a block version of AMRS (B-AMRS) which exploits some topological features of the observed network, such as the block or the core-periphery structure, to design some effective blocking strategy.

We show that the transition kernel of the MRS chain has a mixture representation and, building on \cite{latuszynski2013adaptive}, that the MRS chain is uniformly ergodic. Through a simulation study, we show that MRS overperforms RS for static and multi-layer temporal LS models 
for increasing network order. Moreover, we show that AMRS performs better than MRS in terms of mixing. Finally, we illustrate the efficiency gain of our algorithm on some benchmark applications. 

The structure of this work is as follows. Section \ref{sec:LS_models} introduces LS models. Section \ref{sec:Posterior} describes the standard Gibbs sampler for LS posterior approximation and our novel MRS approach together with its theoretical properties. Section \ref{sec:simulation} provides the results of our simulation study for both static and multi-layer temporal LS models. Section \ref{sec:emp_application} reports the results for two benchmark network datasets. Finally, Section \ref{sec:conclusion} concludes.

\section{Latent Space Models}
\label{sec:LS_models}

\subsection{Simple LS models}
Let $\mathcal{G}=(V, E)$ be a $N$-order graph where  $V \subset \mathbb{N}$ denotes the vertex set, and $E \subset V\times V$ denotes the set of ordered node pairs, that is the edge set. Let $Y$ be the  $N \times N$ adjacency matrix. The $(i,j)$-th entry $y_{ij}$ follows a Bernoulli distribution $y_{ij} \stackrel{ind}{\sim}  \mathcal{B}er(g(\eta_{ij}))$ with parameter $g(\eta_{ij})$ where  $g$ is a link function from the set of the reals $\mathbb{R}$ to the unit interval $[0,1]$,  $\eta_{ij} = \alpha - ||\mathbf{x}_i-\mathbf{x}_j||$ and $||\cdot||$ is a distance between the $d$-dimensional node-specific set of coordinates $\mathbf{x}_i$ for $i = 1, \ldots, N$. Common LS specifications assume a logistic link function $g(\eta)=1/(1+\exp(-\eta))$ and Euclidean distance (\citealp{hoff2002latent,handcock2007model,krivitsky2009representing,friel2016interlocking}). Alternative and less commonly used specifications adopt different distances, such as the squared Euclidean distance (\citealp{gollini2016joint}), the angular distance for hyperbolic spaces (\citealp{Asta2014GeometricNC, smith2019geometry, liu2024bayesian}), the ultrametric distance for ultrametric spaces (\citealp{schweinberger2003settings}),  and link functions, such as the Gaussian link (\citealp{rastelli2016properties}, \citealp{spencer2022faster}).

If the graph is weighted, that is $\mathcal{G} = (V, E, Y)$ with $Y$ a $N\times N$ real-valued weight matrix, then the LS model assumes
$y_{ij} \stackrel{ind}{\sim} f(y_{ij}| g(\eta_{ij}),\kappa)$ $i,j=1,\ldots, N$, $i\neq j$ with parameters $g(\eta_{ij})$ and $\kappa$. The parameter $g(\eta_{ij})$ is driven by node-specific latent features $\eta_{ij} = \alpha  - ||\mathbf{x}_{i}- \mathbf{x}_{j}||$.  When the weights are integer-valued, Poisson, Binomial, or Negative Binomial distributions are assumed (e.g., see \citealp{raftery2017comment}). A truncated-normal or log-normal distribution may be assumed for positive real-valued weights (\citealp{sewell2016latent}, \citealp{egidi2023clustering}).

\subsection{General LS models}

We now define a more general class of LS models, called the multi-layer temporal LS model, which accounts for time variations in the edge weights and also for different weight types.

A temporal weighted graph can be defined as an ordered sequence of graphs that is $\mathcal{G}=\{\mathcal{G}_t\}_{t=1,\ldots, T}$ with $\mathcal{G}_t = (V_t, E_t, Y_t)$. The LS model assumes the $(i,j)$-th entry of $Y_{t}$ satisfies $y_{ijt} \stackrel{ind}{\sim} f(y_{ijt}| g(\eta_{ijt}),\kappa_t)$ $i,j=1,\ldots, N$, $i\neq j$ and $t = 1, \ldots, T$ with parameters $g(\eta_{ijt})$ and $\kappa_t$ which are possibly time-varying. The parameter $g(\eta_{ijt})$ is driven by node- and time-specific latent features $\eta_{ijt} = \alpha_t  - ||\mathbf{x}_{it}- \mathbf{x}_{jt}||$. This formalization considers the discrete-time version of temporal LS. See \citet{rastelli2023continuous} for a continuous-time formalization.

In network analysis, when multiple types of edges are allowed in a temporal graph, the notion of the multi-layer graph is introduced. A temporal multi-layer graph is a sequence of graphs $\mathcal{G}=\{\mathcal{G}_{rqt}\}_{ r,q=1,\ldots, R, t=1,\ldots, T}$ with three indices: the source layer index $r$, the target layer index $q$ and the temporal index $t$ (\citealp{boccaletti2014structure}). A general model with inter- and intra-layer connectivity assumes $\mathcal{G}_{rqt} = (V_{rt}, V_{qt}, E_{rqt}, Y_{rqt})$, where the edge set $E_{rqt}$ defines different types of edges. A typical scenario is given by a node-aligned network, that is, $V_{rt}=V$ for all periods $t$ and layers $r$, and by the absence of inter-layer connectivity, that is $E_{rqt}=\emptyset$ for $r\neq q$. In this case, the temporal multi-layer graph can be defined as $\mathcal{G}=\{\mathcal{G}_{rt}\}_{r=1,\ldots, R, t=1,\ldots, T, }$ where $\mathcal{G}_{rt} = (V, E_{rt}, Y_{rt})$ is the graph which encodes the connectivity of the $r$ layer at time $t$. Different parametrizations of an LS model can be adopted in the case of temporal multi-layer networks. The simplest specifications may assume time- and layer-specific latent coordinates and pooling on both the temporal and the layer dimensions. More complex specifications may use a hierarchical structure to capture layer heterogeneity and induce parameter pooling and dynamic latent processes for modeling the temporal variations. In this work, our results will be presented for the general case of a temporal multi-layer LS model.

\subsection{Prior Choice}
A normal distribution is a standard assumption for the vector of latent coordinates, that is $\mathbf{x}_{irt}\stackrel{iid}{\sim}\mathcal{N}(\mathbf{0},\Sigma)$. The most common choice for the variance-covariance matrix is $\Sigma = \sigma^2I_d$, where the variance parameter $\sigma^2$ is either arbitrarily set or assumed to be inverse-gamma distributed, i.e., $\sigma^2 \sim I\mathcal{G}(a,b)$, with shape parameter and scale parameter $a>0$ and $b>0$ respectively. However, there are cases in which more flexible priors are considered, such as the finite and infinite mixture of Gaussians (e.g., see \citealp{handcock2007model, d2023model}). For what concerns the latent coordinates, it is common to assume random-walk dynamics (e.g., see \citealp{sewell2016latent, friel2016interlocking}), \citet{durante2017bayesian} assumed a Gaussian Process, and \citet{casarin2023media} a Hidden-Markov process.

\section{Posterior Approximation}
\label{sec:Posterior}

In this section, we first provide one possible MCMC scheme via hybrid Gibbs sampling with MH steps for a general version of the multi-layer temporal LS model. Later on, we review RS algorithms and consider the different sampling strategies that have been proposed in the literature. Finally, we introduce the MRS strategy, its adaptive version, AMRS, and its block-adaptive version B-AMRS.

\subsection{Gibbs Sampler}
Consider a dynamic multi-layer network with $R$ layers of $N_r$ nodes observed for $T_r$ time instances.
Let $\mathbf{Y}=(\mathbf{y}_{1},\ldots,\mathbf{y}_{R})$  be the collection of observed network weights $\mathbf{y}_r=(\mathbf{y}_{1r},\ldots,\mathbf{y}_{T_rr})$, $\boldsymbol{\theta} = \left( \boldsymbol{\theta}_1, \ldots, \boldsymbol{\theta}_R \right)$ the collection of parameters $\boldsymbol{\theta}_r 
 = (\alpha_{r1},\ldots,\alpha_{rt},\ldots,\alpha_{rT_r})$ and $\mathbf{X} = (\mathbf{x}_{1}, \ldots, \mathbf{x}_{R})$ the collection of latent coordinates $\mathbf{x}_r=(\mathbf{x}_{1r},\ldots,\mathbf{x}_{ir},\ldots,\mathbf{x}_{N_rr})$, where $\mathbf{x}_{ir}=(\mathbf{x}_{i1r},\ldots,\mathbf{x}_{iT_rr})$ is the $( d \times T_r)$-dimensional matrix of latent coordinates for node $i$. The complete-data likelihood can be written as:
\begin{equation}
f(\mathbf{Y},\mathbf{X} \vert \boldsymbol{\theta})=\prod_{r=1}^{R}\prod_{t=1}^{T_r}\prod_{i=1}^{N_r}\prod_{j = i +1}^{N_r}f(y_{ijrt}|g(\eta_{ijrt}),\kappa_{rt})\pi(\mathbf{X}),
\end{equation}
where $f(y_{ijrt}|g(\eta_{ijrt}),\kappa_{rt})$ is the weight distribution, and
\begin{equation}
\eta_{ijrt} = \alpha_{rt} - \vert\vert \mathbf{x}_{irt}-\mathbf{x}_{jrt}\vert\vert.
\end{equation}

The joint posterior distribution $\pi(\boldsymbol{\theta},\mathbf{X}| \mathbf{Y}) \propto f( \mathbf{Y}| \boldsymbol{\theta}, \mathbf{X})\pi(\boldsymbol{\theta})\pi(\mathbf{X})$ is not tractable. Thus, it is common practice to follow a data augmentation approach and apply Gibbs sampling to approximate the posterior distribution. We approximate the joint posterior distribution by MCMC sampling. Our systematic Gibbs sampling algorithm (hereafter labelled as $GS$) iterates the following steps for each $h$:

\begin{enumerate}
\item Draw $\alpha_{rt}^{(h)}$ from $\pi(\alpha_{rt}| \ldots)$,  for $r = 1, \ldots, R$ and $t = 1, \ldots, T_r$  via MH;
\item Draw $\mathbf{x}_{irt}^{(h)}$ from $\pi(\mathbf{x}_{irt}| \ldots )$, $i=1,\ldots,N_r$ and for $r = 1, \ldots, R$ and $t = 1, \ldots, T_r$ via Adaptive MH (AMH).
\end{enumerate}

A hybrid Gibbs sampler is commonly used when dealing with LS modes, as full conditional distributions are difficult to obtain in closed form. The algorithm involves drawing a candidate value $(\boldsymbol{\theta}^{\ast},\mathbf{X}^{\ast})$ from a proposal distribution $q(\boldsymbol{\theta}^{\ast},\mathbf{X}^{\ast}|\boldsymbol{\theta},\mathbf{X})$, calculating an acceptance probability based on the ratio of the product of likelihood, prior and proposal distribution, i.e. $\alpha((\boldsymbol{\theta}^{\ast},\mathbf{X}^{\ast}),(\boldsymbol{\theta},\mathbf{X}))= \pi(\boldsymbol{\theta}^{\ast},\mathbf{X}^{\ast}|\mathbf{Y})q(\boldsymbol{\theta},\mathbf{X}|\boldsymbol{\theta}^{\ast},\mathbf{X}^{\ast})/(\pi(\boldsymbol{\theta},\mathbf{X}|\mathbf{Y})q(
\boldsymbol{\theta}^{\ast},\mathbf{X}^{\ast}|\boldsymbol{\theta},\mathbf{X}))$, and accepting or rejecting the proposed value with probability $\min\{\alpha((\boldsymbol{\theta}^{\ast},\mathbf{X}^{\ast}),(\boldsymbol{\theta},\mathbf{X})), 1\}$.

Regarding the adaptation of the proposal distribution, we report in Section \ref{alg:haario_amh}, \ref{alg:roberts_amh} and \ref{alg:andrieu_amh} of the Supplementary Materials three well-known AMH algorithms: the adaptive scaling algorithm proposed by \citet{haario2001adaptive},  the global adaptive scaling algorithm proposed by \citet{andrieu2008tutorial} and the incremental scaling algorithm in \citet{roberts2009examples}. The latter algorithms are similar in spirit as they require setting a target acceptance rate $\alpha^{*}$, and deviations of the acceptance rate from the target rate lead to diminishing adjustment of the proposal variance. The fact that the adjustments are diminishing is a condition for the ergodicity of the chains (see \citealp{roberts2007coupling}).

\subsection{Random-Scan Strategies}
\label{subsec:scanstrat}

RS has been primarily designed as a procedure to randomly choose the order of the $N$ parameters to be updated within a Gibbs Sampling algorithm (\citealp{geman1984stochastic, levine2005implementing}). The selection probability of the $i$-th parameter is denoted by $p_i > 0$, and it is commonly assumed that $ p_1 + \ldots + p_N= 1$. In its simplest form, the RS involves assigning to each parameter an equal probability of being updated at each iteration $p_i = N^{-1}$. Algorithm \ref{alg:randomscan} reports a pseudo-code of the Random Scan Gibbs Sampler ($RSG(\mathbf{p})$) that runs for $H$ iterations and $V$ sub-iterations.

\begin{algorithm}
\caption{Random-Scan Gibbs Sampler - $RSG(\mathbf{p})$}
\label{alg:randomscan}

\begin{algorithmic}[1]
\STATE Initialize $\mathbf{X}=\{\mathbf{x}_1, \mathbf{x}_2, \ldots, \mathbf{x}_N\}$ to arbitrary values

\STATE Set the number of iterations $H$, sub-iterations $V$ and probability $p_i$

\FOR{$t = 1$ to $H$}
  \FOR{$v = 1$ to $V$}
    \STATE Sample $i \in \{1,\ldots, N\}$ with probability $p_i$
    \STATE Sample $\mathbf{x}^{(h)}_i$ from $\pi(\mathbf{x}_{i}| \mathbf{X}^{(h-1)}_{-i}, y)$ exactly
  \ENDFOR
\ENDFOR
\end{algorithmic}
\end{algorithm}

Let $(\mathcal{X}, \mathcal{B})$ denote a measurable state space with $\mathcal{X} \subseteq \mathbb{R}^d$, the transition kernel of the $RSG(\mathbf{p})$ defined on  $\mathcal{X} \times \mathcal{B}$ is:
\begin{equation}
P_\mathbf{p}(x, A)=\sum_{i=1}^N p_i \operatorname{P}_i(x, A),
\end{equation}

where  $A \in \mathcal{B}(\mathcal{X})$ denoting a measurable set and $\operatorname{P}_i(x, A)$ is the kernel of the Gibbs step for node $i$ that involves updating $\mathbf{x}_i$ from $\pi(\mathbf{x}_{i}| \mathbf{X}_{-i}, y)$.

The Adaptive RS algorithm is commonly described as an RS strategy for which the set of selection probabilities $\mathbf{p}^{(h)}=\{p^{(h)}_1, p^{(h)}_2, \ldots,p^{(h)}_N\}$ at iteration $h$ are determined by some decision rule $R(\mathbf{p}^{(h)}|\mathbf{X}^{(1:h-1)})$. As in Algorithm \ref{alg:randomscan}, the selection probabilities are such that $p_1^{(h)}+ \ldots + p_{N}^{(h)} = 1$ for each $h$.

Algorithm \ref{alg:adrandomscan} reports the pseudo-code of a general Adaptive RS Gibbs Sampler ($ARSG(\mathbf{p})$). In the algorithm, the selection probabilities get updated at each iteration $h = 1, \ldots, H$ while they remain constant along each sub-iteration $v^{(h)} = 1, \ldots, V$. The choice of the number of sub-iterations $V$ is delegated to users, who should decide how often they require the update of the selection probabilities.

\begin{algorithm}
\caption{General Adaptive Random Scan Gibbs Sampler - $ARSG(\mathbf{p})$}
\label{alg:adrandomscan}

\begin{algorithmic}[1]
\STATE Initialize $\mathbf{X}=\{\mathbf{x}_1, \mathbf{x}_2, \ldots, \mathbf{x}_N\}$ to arbitrary values

\STATE Set the number of iterations $H$ and sub-iteations $V$

\FOR{$h = 1$ to $H$}
   \STATE $\mathbf{p}^{(h)} \leftarrow R(\mathbf{p}^{(h)}|\mathbf{X}^{(1:h-1)})$

  \FOR{$v = 1$ to $V$}
    \STATE Sample $i \in \{1,\ldots, N\}$ with probability $p^{(h)}_i$
    \STATE Sample $\mathbf{x}^{(h)}_i$ from $\pi(\mathbf{x}_{i}| \mathbf{X}^{(h-1)}_{-i}, y)$ exactly
  \ENDFOR
\ENDFOR
\end{algorithmic}
\end{algorithm}

Several works provided alternative decision rules aimed at exploring more frequently parameters that exhibit higher variability. \citet{levine2006optimizing} suggest the use of the min-max random scan, which consists of finding the min-max solution $\mathbf{p}^{*}$ with respect to the expected loss $R(\mathbf{p}, g)$  where $g$ is an appropriate function in $L^2(\pi)$. As reported also in \citet{levine2005implementing},  $R(\mathbf{p}, g)$  can be chosen according to convergence rate and asymptotic variance considerations. While the methodology appears appealing, it might be challenging to implement an on-the-fly optimization when there is no analytical solution for the risk function. However, the authors suggest that relying on Gaussian approximation might be a viable option, which requires some analytical tractability of the posterior.

\citet{latuszynski2013adaptive} propose alternative versions of the RS Gibbs and Metropolis-within-Gibbs samplers and suggest that not only the selection probabilities but also the proposal distribution of the MH step can be adapted as a standard practice. Finally, \citet{chimisov2018adapting} introduce a general-purpose RS Gibbs sampler or Metropolis-within-Gibbs for which the selection probabilities are chosen to optimize the pseudo-spectral gap of the chains.

\subsection{An Adaptive Multiple Random-Scan for LS models}
Hereafter, we propose a novel Multiple RS scheme (hereafter \emph{MRSG}$(\mathbf{q})$), its adaptive version (hereafter \emph{AMRSG}$(\mathbf{q})$), and its block-adaptive version (hereafter \emph{B-AMRSG}$(\mathbf{q})$).

At every iteration, \emph{MRSG}$(\mathbf{q})$ selects the components to update. This is done by drawing $b_i \in \{0,1\}$ from a Bernoulli distribution with parameter $q_i$, for $i = 1, \ldots, N$ until $b_1 + \ldots + b_N > 0$. This is equivalent to draw the random indices $s_j$ in the random set $\mathfrak{I} = \{i = 1, \ldots, N, \, s.t. \,\, b_i = 1 \}$ with cardinality $M = Card(\mathfrak{I} )$. The joint distribution of $M$ and $(s_1, \ldots, s_M)$ is denoted by $g(\mathbf{q})$.
We exploit the adaptation in the MH proposal to adapt the probability $q_i$ of selecting a given node $i$ for the update. We report the algorithmic implementation of the \emph{MRSG}$(\mathbf{q})$ in Algorithm \ref{alg:mrsg}.

\emph{MRSG}$(\mathbf{q})$ randomizes not only which latent positions get updated, but also the number of updates at each iteration. For this reason, the probabilities $q_i$ do not add up to one. We notice how drawing a subset of latent coordinates rather than just one latent coordinate at each iteration provides better behavior for the chains, especially for cases in which on-the-fly re-centering of the latent coordinates is implemented (see \citealp[p. 396]{gelman1995bayesian}, \citealp{keefe2018formal}). We prevent the case in which no variable is selected by re-drawing until at least one set of coordinates is updated. 

As an adaptive rule, we follow a simple heuristic according to which latent positions with acceptance rates lower than the target rate should be sampled more often, while latent positions with an acceptance rate higher than a target should be sampled less often. We link the difference between the actual acceptance rate and the target rate via a strictly decreasing function $\varphi$ such that $\varphi: \mathbb{R} \rightarrow [0,1]$.

In this work, we narrow our scope by considering the use of a flipped logistic function, and we implement the probability adaptation procedure every $u$ iterations. The specification is the following:
\begin{equation}
\label{eq3:update_prob}
 q_i^{(h)} = 1/\left(1 + \exp\{\Bar{a}_i^{(h)} - \alpha^* + c\} \right),
\end{equation}
where $\Bar{a}_i^{(h)} = \frac{1}{u}\sum_{i= h-u +1}^{h}a_i^{(h)}$ denotes the average instantaneous acceptance rate from the last probability-adaptation step $h- u + 1$ until iteration $h$ for the node $i$, $\alpha^*$ denotes the target acceptance rate, and $c \in \mathbb{R}$ denotes a shift parameter to be chosen. The choice of $u$ allows deciding whether adaptation should be performed more or less often. Algorithm \ref{alg:adrandomscan_as} reports our \emph{AMRSG}$(\mathbf{q})$,  in which individual adaptation of the selection probability for each latent position is implemented. Although Algorithm \ref{alg:adrandomscan_as} shares the same time complexity, i.e., $\mathcal{O}(N^2)$, of many other MCMC algorithms for LS models, it allows reducing the computational cost by a factor by limiting the number of nodes for which the latent coordinates get updated at each iteration.

The uniform ergodicity of $RSG(\mathbf{p})$ is given in \citealp[][Th. 4]{latuszynski2013adaptive}. In the following, we state the same property for the multi-step random scan sampler with fixed selection probabilities, \emph{MRSG}$(\mathbf{q})$, given in Algorithm \ref{alg:adrandomscan}.

\begin{proposition}
\label{prop:rsg}
    Let $\mathbf{q} \in [\epsilon, 1]^N$ with $\epsilon > 0$. If the systematic scan Gibbs sampler ($GS$) is uniformly ergodic, so is \emph{MRSG}$(\mathbf{q})$.
\end{proposition}

\begin{proof}
    See Appendix A.
\end{proof}

Under the assumptions posed by \citet{latuszynski2013adaptive} on the selection probabilities and following the results of the previous proposition, we show that the adaptive version of the multiple random scan introduced in Algorithm \ref{alg:adrandomscan_as} and denoted by \emph{AMRSG}$(\mathbf{q})$ is ergodic. Let $\pi^{(h)}\left(\mathbf{x}^{(0)}, \mathbf{q}^{(0)}\right)$ denote the distribution at iteration $h$ induced by Algorithm \ref{alg:adrandomscan_as} with starting values $\mathbf{x}^{(0)}$ and $\mathbf{q}^{(0)}$, $\pi$ the stationary distribution of the Markov chain, and $||\cdot||_{TV}$ the total variation norm.
 
\begin{proposition}
Let the selection probabilities $\mathbf{q}^{(h)} \in [\epsilon, 1]^N$ for all $h$ and $\epsilon > 0$. Assume that:

\begin{itemize}
    \item[a)] $\left|\mathbf{q}^{(h)}-\mathbf{q}^{(h-1)}\right| \rightarrow 0$ in probability for fixed starting values $\mathbf{x}^{(0)} \in \mathcal{X}$ and $\mathbf{q}^{(0)}  \in [\epsilon, 1]^N$.
    \item[b)] there exists $\mathbf{q} \in [\epsilon, 1]^N$ s.t. $\operatorname{RSG}(\mathbf{q})$ is uniformly ergodic.
\end{itemize}
Then \emph{ARSG}($\mathbf{q}$) is uniformly ergodic, that is:
$$||\pi^{(h)}\left(\mathbf{x}^{(0)}, \mathbf{q}^{(0)}\right) - \pi ||_{TV} \rightarrow 0 \quad \text { as }  h \rightarrow \infty.$$
\end{proposition}

\begin{proof}
    See Appendix A.
\end{proof}

The assumption b) can be verified by exploiting Proposition \ref{prop:rsg}. For what concerns assumption a), \citet{latuszynski2013adaptive} argue that most adaptive Gibbs samplers will satisfy the assumption provided one finds a deterministic sequence $b^{(h)} \rightarrow 0$ such that $|q^{(h)} - q^{(h -1)}| \leq b^{(h)}$. 

 \begin{remark}
 To ensure that assumption a) of Proposition 3.3.2 holds, one can assume the update of the selection probabilities:
 
\begin{equation*}
    \mathbf{q}^{(h)}=\mathbf{q}^{(h-1)}(1-b^{(h)})+Db^{(h)},
\end{equation*}
with $D=\mathbf{q}^*-\mathbf{q}^{(h-1)}$ and where $\mathbf{q}^*$ is the proposed update in Equation \ref{eq3:update_prob} and $b^{(h)} \in [0,1]$ is a deterministic sequence such that $b^{(h)} \rightarrow 0$.
\end{remark}

Algorithm \ref{alg:adrandomscan_block} reports an alternative sampling method with block adaptation (\emph{B-AMRSG}$(\mathbf{q})$). This algorithm exploits the clear topological features of the network that can be detected by some statistics. 
For example, several networks exhibit a core-periphery structure (e.g., see \citealp{csermely2013structure}), and many off-the-shelf algorithms are available for the detection of core and periphery nodes (see \citealp{ma2015rich} for an example).

In the blocks version of the algorithm, nodes are split into K nonoverlapping partitions or blocks with cardinality $N_k$ such that $N_1 + \ldots + N_K = N$ and the coordinates get updated according to the draw from a Bernoulli distribution with parameter $q_k$. The probability according to which we updated the latent coordinates of the nodes belonging to partition $k$ is defined as in \ref{eq3:update_prob}.

In this algorithm, the updated probabilities $q_k^{(h)}$ are normalized, i.e. $\mathbf{q}^{(h)}/||\mathbf{q}^{(h)}||_1 $, where $\mathbf{q}^{(h)} = (q_1^{(h)},\ldots, q_K^{(h)})$.

\begin{algorithm}
\caption{Multiple Random Scan Gibbs Sampler - \emph{MRSG}$(\mathbf{q})$)}
\label{alg:mrsg}

\begin{algorithmic}[1]
\STATE Initialize $\mathbf{X}=\{\mathbf{x}_1, \mathbf{x}_2, \ldots, \mathbf{x}_N\}$ to arbitrary values
\STATE Initialize $\mathbf{q}$

\STATE Choose a target acceptance rate $\alpha^*$ and set AMH parameters, e.g. as in Algorithms \ref{alg:haario_amh}, \ref{alg:roberts_amh}, \ref{alg:andrieu_amh}
\STATE Set the number of iterations $H$

\FOR{$h = 1$ to $H$}
 \STATE Sample the random indices $(M, s_1, \ldots, s_M) \sim g(\mathbf{q}^{(h-1)})$ s.t. $1\leq M \leq N$ and $s_j \in \{1, \ldots, N\}$ are distinct with $j \in 1, \ldots, M$.
  \FOR{$i = 1$ to $M$}
      \STATE Sample $\mathbf{x}_{s_i}$ from $\pi(\mathbf{x}_{s_i}| \mathbf{X}^{(h-1)}_{-s_i}, y)$ via AMH with $\mathbf{X}^{(h-1)}_{-s_i}=\mathbf{X}^{(h-1)}/\{\mathbf{x}^{(h-1)}\}$ 
 \ENDFOR
\ENDFOR
\end{algorithmic}
\end{algorithm}

\begin{algorithm}
\caption{Adaptive Multiple Random Scan Gibbs Sampler - \emph{AMRSG}$(\mathbf{q})$}
\label{alg:adrandomscan_as}

\begin{algorithmic}[1]
\STATE Initialize $\mathbf{X}=\{\mathbf{x}_1, \mathbf{x}_2, \ldots, \mathbf{x}_N\}$ to arbitrary values
\STATE Choose a target acceptance rate $\alpha^*$ \\ (and set AMH parameters, e.g. as in Algorithms \ref{alg:haario_amh}, \ref{alg:roberts_amh}, \ref{alg:andrieu_amh})
\STATE Set the number of iterations $H$, the probability-adaptation step $u$, and the constant $c \in \mathbb{R}$

\FOR{$h = 1$ to $H$}
 \STATE Sample the random indices $(M, s_1, \ldots, s_M) \sim g(\mathbf{q}^{(h-1)})$ s.t. $1\leq M \leq N$ and $s_j \in \{1, \ldots, N\}$ are distinct with $j \in 1, \ldots, M$.
  \FOR{$i = 1$ to $M$}
      \STATE Sample $\mathbf{x}_{s_i}$ from $\pi(\mathbf{x}_{s_i}| \mathbf{x}^{(h-1)}_{-s_i}, y)$ via AMH with $\mathbf{X}^{(h-1)}_{-s_i}=\mathbf{X}^{(h-1)}/\{\mathbf{x}_{(h-1)}\}$ 
      \STATE Store $ a_{s_i}^{(h)} = \min\{\alpha(\mathbf{X}^{(h-1)}, \Tilde{\mathbf{X}}), 1\}$ where $\Tilde{\mathbf{X}} = (\mathbf{x}^{(h-1)}_1\!\!,\ldots,\mathbf{x}^{(h-1)}_{s_i -1},\mathbf{x}_{s_i}, \mathbf{x}^{(h-1)}_{s_i +1}\!, \ldots, \mathbf{x}^{(h-1)}_{N} )$

  \IF{$ [\frac{h}{u}]= 0$}
    \STATE Set $\Bar{a}_{s_i}^{(h)} = \frac{1}{u}\sum_{j= h-u +1}^{h}a_{s_i}^{(j)}$
    \STATE Set $q_{s_i}^{(h)} = 1/\left(1 + \exp\{\Bar{a}_{s_i}^{(h)} - \alpha^* + c\} \right)$
  \ENDIF
 \ENDFOR
\ENDFOR
\end{algorithmic}
\end{algorithm}

\begin{algorithm}
\caption{Block Adaptive Multiple Random Scan Gibbs Sampler - \emph{B-AMRSG}$(\mathbf{q})$}
\label{alg:adrandomscan_block}

\begin{algorithmic}[1]
\STATE Initialize $\mathbf{x}_1, \mathbf{x}_2, \ldots, \mathbf{x}_N$ to arbitrary values
\STATE Choose a target acceptance rate $\alpha^*$ and set AMH parameters, e.g. as in Algorithms \ref{alg:haario_amh}, \ref{alg:roberts_amh}, \ref{alg:andrieu_amh}
\STATE Set a partition with $K$ non-intersecting elements of the node set $\{1,\ldots,N\}$ with elements $\mathcal{N}_1,\ldots,\mathcal{N}_K$, and denote with ${\mathbf{X}_1, \mathbf{X}_2, \ldots, \mathbf{X}_K}$ the corresponding sets of coordinates, that is $\mathbf{X}_j= \{\mathbf{x}_{\ell}\}_{\ell\in\mathcal{N}_j}$ with cardinality $N_j$
\STATE Set the number of iterations $H$, the probability-adaptation step $u$, and the constant $c \in \mathbb{R}$

\FOR{$h = 1$ to $H$}
    \STATE Sample the random indices $(M, s_1, \ldots, s_M) \sim g(\Tilde{\mathbf{q}}^{(h-1)})$ s.t. $1\leq M \leq K$ and $s_j \in \{1, \ldots, K\}$ are distinct with $j \in 1, \ldots, M$.

  \FOR{$ k = 1$ to $M$}
     \FOR{$i\in \mathcal{N}_k$}
      \STATE Sample $\mathbf{x}_{i}$ from $\pi(\mathbf{x}_{i}| \mathbf{X}^{(h-1)}_{-i
      }, y)$ via Adaptive Metropolis-Hastings, where $\mathbf{X}^{(h-1)}_{-i}=\mathbf{X}^{(h-1)}/\{\mathbf{x}_{i}^{(h-1)}\}$
       \STATE Store $ a_{i}^{(h)} = \min\{\alpha(\mathbf{X}^{(h-1)}, \Tilde{\mathbf{X}}), 1\}$ where $\Tilde{\mathbf{X}}= \left(\mathbf{X}^{(h-1)}/ \{\mathbf{X}_k^{(h-1)}\}\right)\cup \{\Tilde{\mathbf{X}}_k\}$, \\$\Tilde{\mathbf{X}}_k= \left(\mathbf{X}_k^{(h-1)}/ \{\mathbf{x}_{i}^{(h-1)}\}\right)\cup \{\mathbf{x}_{i
       }\}$
       \ENDFOR
      \STATE Set $a_{{s_k}}^{(h)} = \frac{1}{N_{s_k}}\sum_{i\in\mathcal{N}_k} a_{i}^{(h)}$
  \IF{$ [\frac{h}{u}]= 0$ }
    \STATE Set $\Bar{a}_{s_k}^{(h)} = \frac{1}{u}\sum_{j= h-u +1}^{h}a_{s_k}^{(j)}$
    \STATE Set $q_{s_k}^{(h)} = 1/\left(1 + \exp\{\Bar{a}_{s_k}^{(h)} - \alpha^* + c\} \right)$
  \ENDIF
    \STATE Set  $\Tilde{\mathbf{q}}^{(h)} = \mathbf{q}^{(h)}/||\mathbf{q}^{(h)}||_1 $ 
    \ENDFOR
\ENDFOR
\end{algorithmic}
\end{algorithm}

\section{Simulation Study}
\label{sec:simulation}

\subsection{Synthetic Datasets}
\begin{figure}[htbp]
    \centering
    \begin{tabular}{cc}
        \begin{subfigure}[b]{0.45\textwidth}
            \centering
            \includegraphics[width=\textwidth]{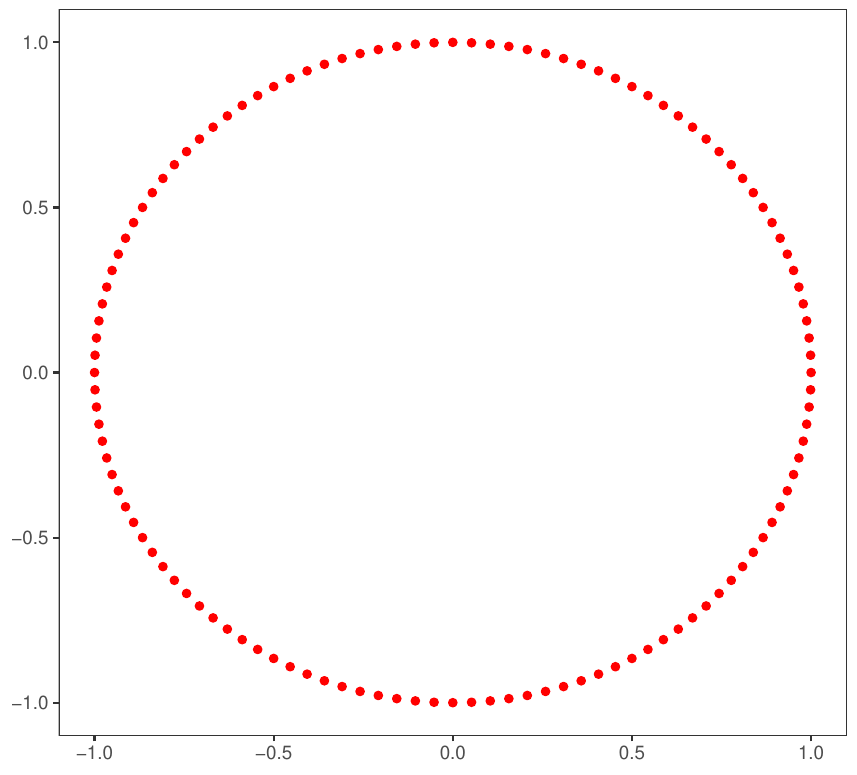}
            \caption{\begin{tabular}{c}Circular Layout.\\ Latent Coordinates (red dots).\end{tabular}}
            \label{fig:3circle_1}
        \end{subfigure}
        &
        \begin{subfigure}[b]{0.45\textwidth}
            \centering
            \includegraphics[width=\textwidth]{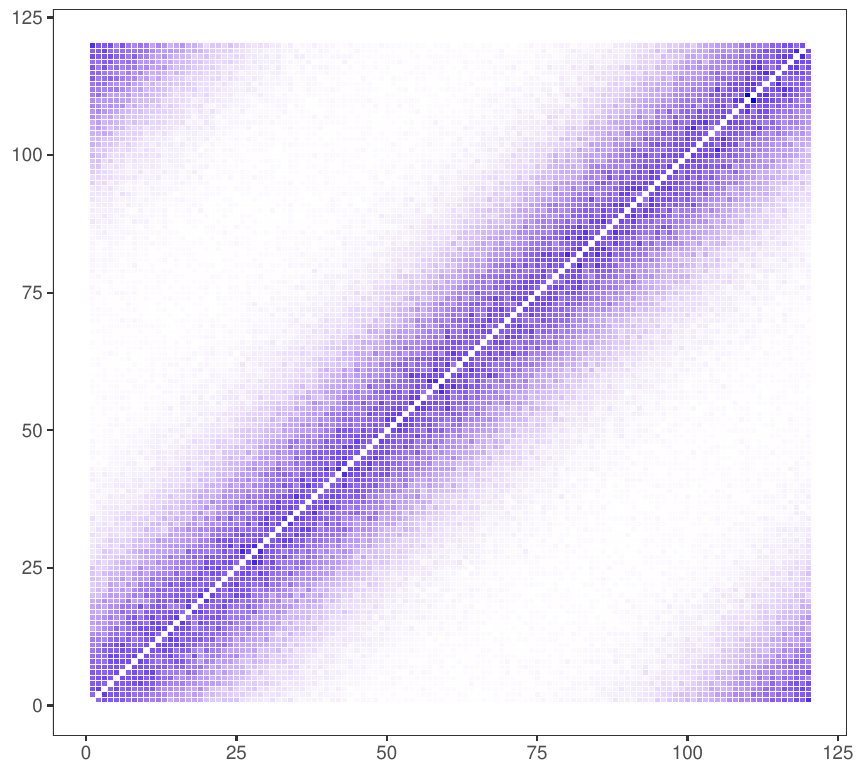}
            \caption{\begin{tabular}{c}  Circular Layout.\\ Poisson Model realization.\end{tabular}}
            \label{fig:3circle_2}
        \end{subfigure}\\
        
        \begin{subfigure}[b]{0.45\textwidth}
            \centering
            \includegraphics[width=\textwidth]{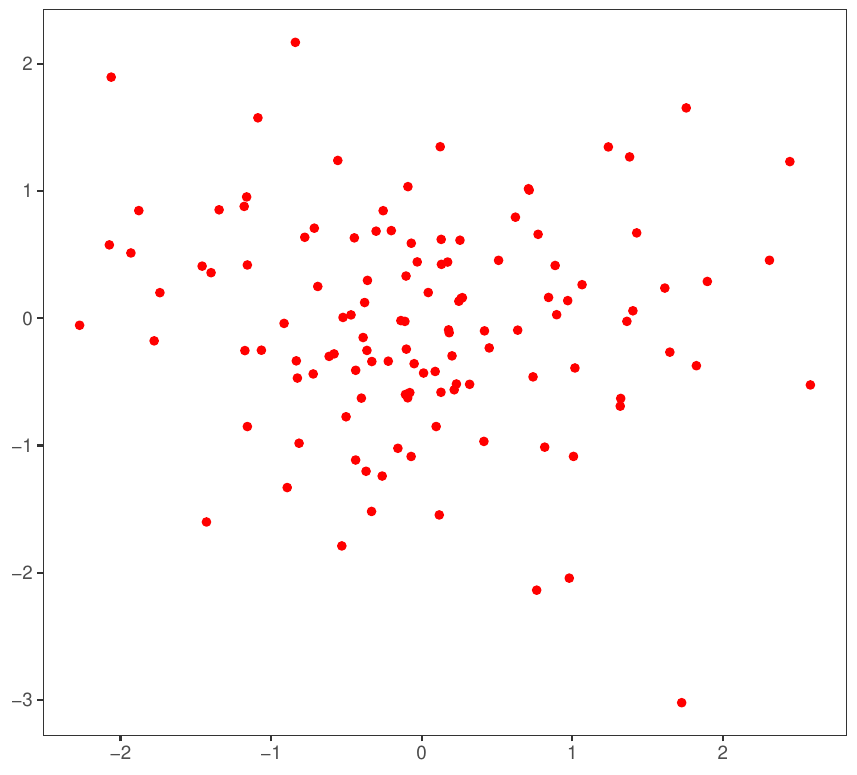}
            \caption{\begin{tabular}{c} Random Layout.\\ Latent Coordinates (red dots).\end{tabular}}
            \label{fig:3random_1}
        \end{subfigure}
        &
        \begin{subfigure}[b]{0.45\textwidth}
            \centering
            \includegraphics[width=\textwidth]{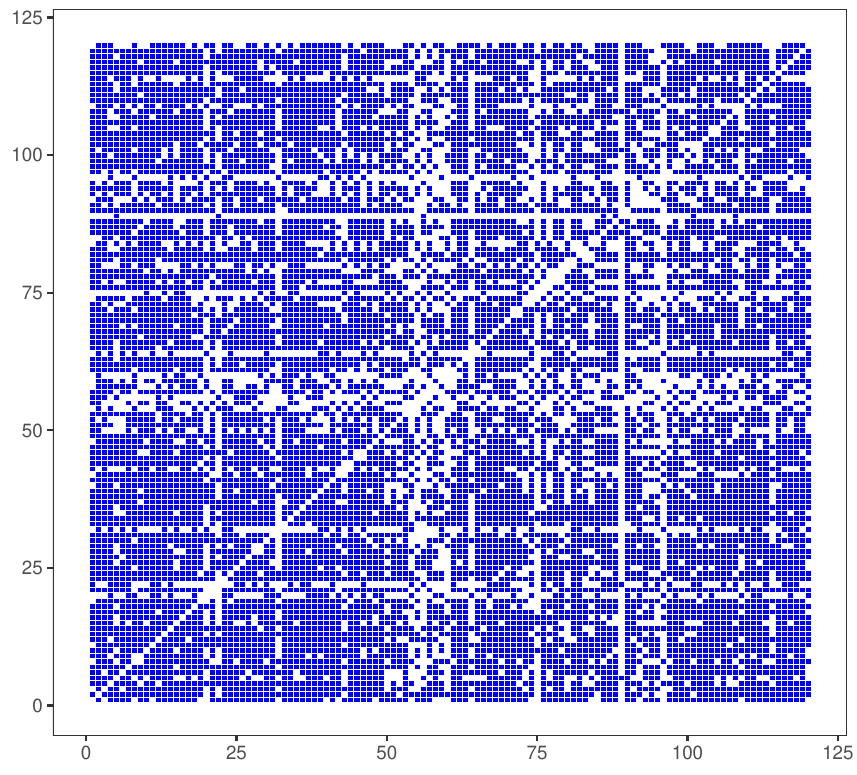}
            \caption{\begin{tabular}{c}  Random Layout.\\ Bernoulli Model realization.\end{tabular}}
            \label{fig:3random_2}
        \end{subfigure}
    \end{tabular}
    
    \caption{\textbf{Static Synthetic Networks:} Representation of the nodes on the latent space (left) and weighted adjacency matrix of the simulated network (right), color gradient proportional to the weight of the edge for circular (top) and random (bottom) graphs. The network has 120 nodes, and the LS model has parameters $\alpha = 5$ and $d=2$.}
    \label{fig:3layout}
\end{figure}

\begin{figure}[htbp]
    \centering

    \centering
            \includegraphics[width=\textwidth]{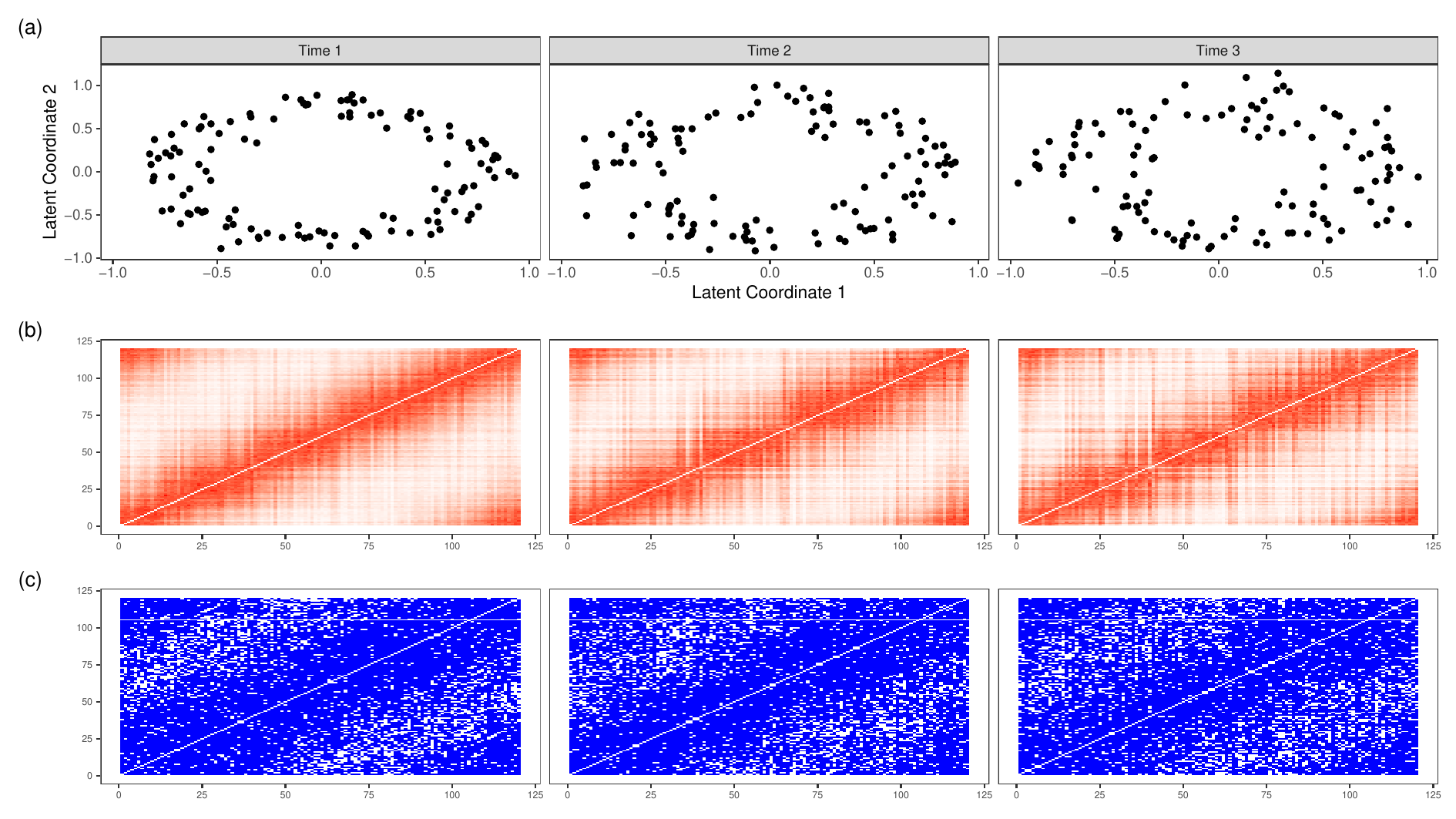}
           
    \caption{\textbf{Multi-layer Temporal Synthetic Network:} Panel a) reports the latent coordinates at different time instances. Panel b) reports the adjacency matrix representation of the Poisson weighted network. Panel c reports the adjacency matrix representation of the Bernoulli binary network. The network has 120 nodes, $\alpha = 5$ and $d=2$.}
    \label{fig:3layoutMT}
\end{figure}

To assess the efficiency of our algorithm, we perform a simulation study on three synthetic networks: two static and one multi-layer temporal.
We generate the synthetic datasets from a starting configuration of the nodes on the latent space.
We restrict our analysis to the case in which the latent space is a plane ($d=2$). For the static setup, we focus on two possible layouts of the nodes on the plane: (i) nodes lie on a circle of unit radius centered at zero; (ii) nodes are randomly disposed on the latent space according to their prior, i.e., $\mathbf{x}_{i}\stackrel{iid}{\sim}\mathcal{N}(\mathbf{0},\Sigma)$ and $\Sigma = I_d$. Given a set of $N$ latent coordinates, we generate an adjacency matrix $A_{N \times N}$ from the data-generating process (DGP) of a general LS model. In particular, we focus on the case $y_{ij} \stackrel{ind}{\sim} \mathcal{P}oi(y_{ij}| \exp(\eta_{ij}))$ for the circular layout and $y_{ij} \stackrel{ind}{\sim} \mathcal{B}er\left(y_{ij}| \frac{1}{1+\exp(-\eta_{ij})}\right)$ for the random layout with $\eta_{ij} = \alpha - \vert\vert \mathbf{x}_{i}-\mathbf{x}_{j}\vert\vert^2$ where $\vert\vert \mathbf{x}_{i}-\mathbf{x}_{j}\vert\vert^2$ denotes the squared Euclidean distance and $\alpha = 5$. Figure \ref{fig:3layout} reports the two different static layouts of the nodes on the plain (circular and random) and the corresponding adjacency matrices under the aforementioned DGP. For the multi-layer temporal setup, we consider $y_{ijrt} \stackrel{ind}{\sim} f_r(y_{ijrt}| g_r(\eta_{ijt}),\kappa_t)$ $r=1,2$ and $t =1,\ldots 3$ for which $f_r \in \{\mathcal{P}oi, \mathcal{B}er\}$, $g_r(\eta_{ijrt}) \in \{exp(\eta_{ijrt}), (1+exp(-\eta_{ijt}))^{-1}\}$ and $\eta_{ijt} =  \alpha_{rt} - \vert\vert \mathbf{x}_{it}-\mathbf{x}_{jt}\vert\vert^2$ with $\alpha_{rt} \in \{6,3\}_{t=1}^{3}$ and $\mathbf{x}_{it} =  \mathbf{x}_{it-1} + \epsilon_{it}$. We assume a circular setup for $\mathbf{x}_{0}$ and $\epsilon_{it} \sim \mathcal{N}(0, \sigma^2_\epsilon)$ with $\sigma^2_\epsilon = 0.01$. Figure \ref{fig:3layoutMT} illustrates a multi-layer temporal network.

\subsection{A Comparison}
\label{subsec:algorithms}
The algorithms considered for comparison are the following. The standard systematic Gibbs algorithm ($GS$), the individual-update equal selection-probability Multiple RS with probability  $q_i \in \{0.25, 0.5\}$ for all $i = 1, \ldots, N$ (labelled as, \emph{MRSG}$_{0.25}$ and \emph{MRSG}$_{0.5}$), the Block Multiple RS with $K \in \{2,4\}$ and $q_k = 1/K$ for all $k = 1, \ldots, K$ (\emph{B-MRSG}$_{2}$ and \emph{B-MRSG}$_{4}$), the individual-update adaptive selection-probability Multiple RS (\emph{AMRSG}) and the block adaptive Multiple RS with $K \in \{2, 4\}$ (\emph{B-AMRSG}$_{2}$ and \emph{B-AMRSG}$_{4}$). The AMH step we implement is the one suggested by \citet{andrieu2008tutorial} and the target acceptance rate $\alpha^{*}$ is set equal to 0.234 in each adaptive algorithm. We also set the shift parameter $c = 0$.

\subsection{Simulation Design and Assessment Metrics}

We run 250 parallel batches of each algorithm on a computing system with 64 cores and 728GB of RAM. Each algorithm has been run for 30'000 iterations. The algorithmic implementations are written in C++ and can be invoked as functions in R through the \texttt{Rcpp} package.

We compare the performance of the algorithms across different domains: estimation error, precision, MCMC chain mixing, and running time. As an indicator of estimation error, we adopt the Mean Squared Error (MSE), computed for each draw of the latent coordinates and averaged across latent coordinates. As an indicator of precision, we compute the variance of the chains. An estimate of Effective Sample Size (ESS) defined as $ESS =\frac{N}{1+2 \sum_{t=1}^{\infty} \rho_t}$ is instead used to determine how well an MCMC chain is mixing (\citealp{gelman1995bayesian}). The estimate we rely on is the one provided by the R package \texttt{LaplacesDemon} (\citealp{hall2008laplacesdemon}), i.e. $\widehat{ESS} =\frac{N}{1+2 \sum_{t=1}^{K} \hat{\rho}_t}$ where K is the first lag for which $|\hat{\rho}_t| < 0.05$. Finally, we compute the running time (expressed in seconds) of each algorithm to determine which is faster. The simulation results in the next section will be presented without applying burn-in and thinning. This is done as one may want to penalize algorithms with slower convergence.

\subsection{Static Setup: Results}
For the sake of the exposition, we report the results for the circular-layout network here, while the random layout results are reported in the Supplementary Materials, Section \ref{apx:rand}.

\begin{figure}[!htb]
  \centering
   \includegraphics[width= \textwidth]{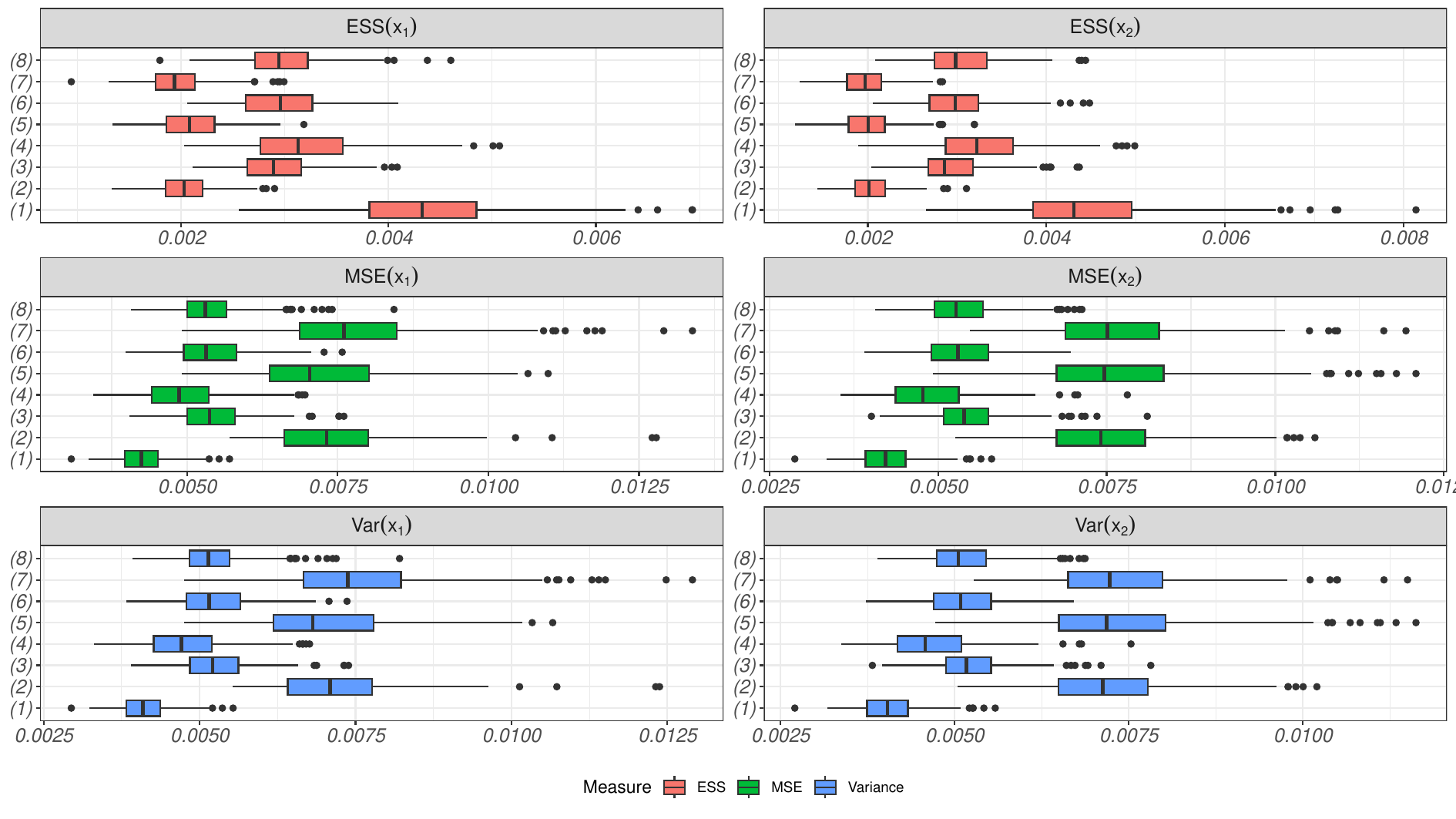}
   \caption{\textbf{Metrics - No Burn-in and Thinning on Circular-Layout Network:} Comparison between the competing algortithms. The boxplots report metrics comparison for 250 runs of the algorithms in \ref{subsec:algorithms}. The reported metrics are the Effective Sample Size (ESS) as a proportion of the overall sample, the Mean Squared Error (MSE) compared to the true value of the latent coordinates, and the variance of the chains. The metrics are averaged across nodes for each latent coordinate $\mathbf{x}_1$ (left) and $\mathbf{x}_2$ (right). The number of iterations has been set to 30'000. The adaptive selection probabilities get updated every 100 iterations. Legend: (1) GS, (2) \emph{MRSG}$_{0.25}$, (3) \emph{MRSG}$_{0.5}$, (4) \emph{AMRSG}, (5) \emph{B-MRSG}$_4$, (6) \emph{B-MRSG}$_2$, (7)  \emph{B-AMRSG}$_4$, and (8) \emph{B-AMRSG}$_2$.}
 \label{fig:3metrics_noburnin}
\end{figure}

Figure \ref{fig:3metrics_noburnin} provides a comparison across the aforementioned algorithms in terms of $\widehat{ESS}$, MSE, and Variance before burnin-in and thinning. $GS$ exhibits the best mixing, i.e. larger $\widehat{ESS}$, followed by \emph{B-AMRSG}$_2$, \emph{B-MRSG}$_2$, \emph{AMRSG}, and \emph{MRSG}$_{0.5}$, while \emph{B-AMRSG}$_4$, \emph{B-MRSG}$_4$ and \emph{MRSG}$_{0.25}$ exhibit the worst mixing. MSE and variance show a similar pattern across algorithms: $GS$ has the lowest MSE (Variance) followed by \emph{B-AMRSG}$_2$, \emph{B-MRSG}$_2$, \emph{AMRSG}, and \emph{MRSG}$_{0.5}$, while \emph{B-AMRSG}$_4$, \emph{B-MRSG}$_4$, and \emph{MRSG}$_{0.25}$ exhibit the highest. As one can expect, reducing the number of nodes updated at each iteration deteriorates the efficiency of the sampler. In our experiments, updating each node with $50$\% probability has a negligible impact on the efficiency.
Moreover, \emph{AMRSG} exhibits better results than \emph{MRSG}$_{0.5}$ (its direct competitor), which implies that an adaptive selection probability rule can pay off compared to the non-adaptive equal probability selection scheme. For what concerns group updating, the adaptive selection probability rule does not exhibit any relevant difference w.r.t. the equal-probability counterpart in terms of estimation error, precision, and mixing. Figure \ref{fig:metrics_few_nodes} in Appendix \ref{apx:rand} reports a similar exercise in which the algorithms run for 5'000 iterations. We do not notice any substantial difference with the aforementioned results in terms of convergence for all the considered algorithms. Figure \ref{fig:ks_diagnostic} in Appendix \ref{apx:rand} reports an alternative convergence assessment of the chains based on a sequence of Kolmogorov-Smirnov tests (see, \citealp{robert1999monte}, pp. 466-470). This assessment confirms that \emph{AMRSG}, i.e., our adaptive RS, converges slightly faster than its direct competitor, \emph{MRSG}$_{0.5}$, the RS with fixed selection probabilities.

\begin{figure}[!htb]
  \centering
   \includegraphics[width= 0.9\textwidth]{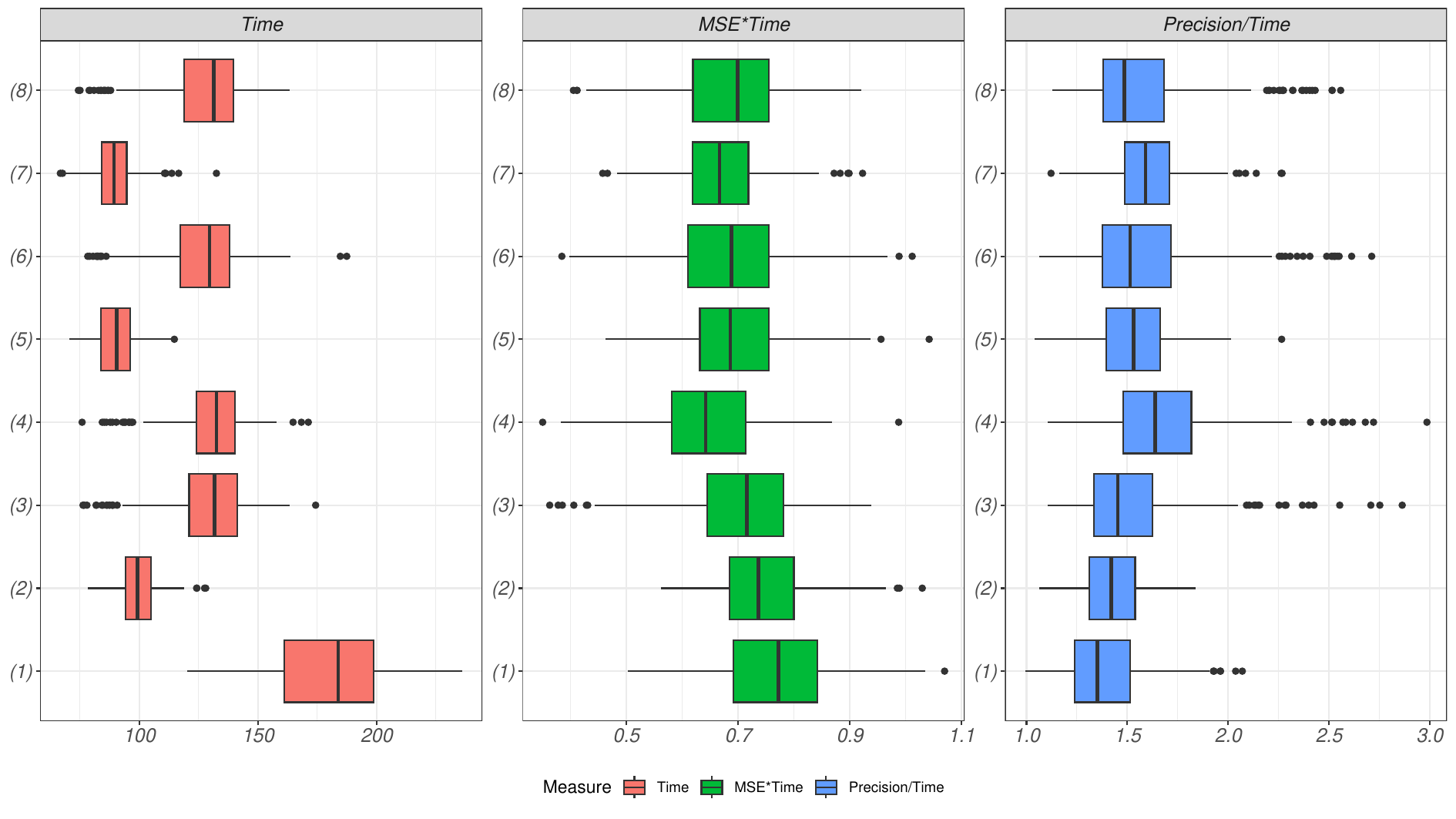}
   \caption{\textbf{Timing Circular Layout:} Comparison between the competing algorithms in terms of computing time in seconds, $MSE*Time$, and precision-to-time ratio, where precision is computed as the inverse of the variance. The boxplots report metrics comparison for 250 runs of the algorithms in \ref{subsec:algorithms}. The metrics are averaged across nodes and coordinates. The adaptive selection probabilities get updated every 100 iterations. Legend: (1) GS, (2) \emph{MRSG}$_{0.25}$, (3) \emph{MRSG}$_{0.5}$, (4) \emph{AMRSG}, (5) \emph{B-MRSG}$_4$, (6) \emph{B-MRSG}$_2$, (7)  \emph{B-AMRSG}$_4$, and (8) \emph{B-AMRSG}$_2$.}
 \label{fig:3timing}
\end{figure}

Panel 1 in Figure \ref{fig:3timing} provides a comparison accounting for the algorithm computing times. As one can expect, reducing the number of positions at each iteration leads to a reduced computing time. $GS$ is the slowest algorithm and requires $\sim$ 175 seconds to run. \emph{AMRSG}, \emph{MRSG}$_{0.5}$, \emph{B-AMRSG}$_2$ and \emph{B-MRSG}$_2$ require $\sim$ 125 seconds (approx. 30\% faster), while  \emph{MRSG}$_{0.25}$,  \emph{B-AMRSG}$_4$ and \emph{B-MRSG}$_4$ require $\sim$ 80 seconds (approx. 50\% faster). The worst-performing algorithms in terms of $\widehat{ESS}$, $MSE$, and variance are now the best in computing time since there is a trade-off between accuracy and computing time, For this reason, Panel 3 in Figure \ref{fig:3timing} reports the metric $MSE*Time$ and the precision-to-time ratio. According to both criteria, \emph{AMRSG} and  \emph{B-AMRSG}$_4$ seem to be slightly preferred.

\begin{figure}[!htb]
  \centering
   \includegraphics[width= 0.95\textwidth]{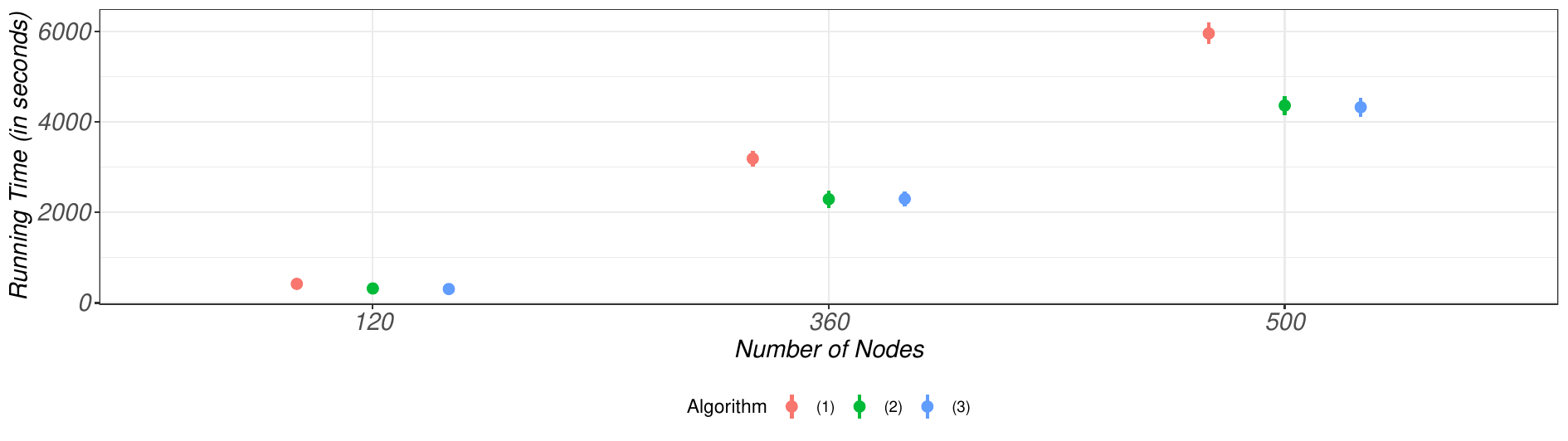}
   \caption{\textbf{Running Time at increasing network size - Circular Layout:} Comparison of the running time of GS (1), \emph{AMRSG} (2), \emph{MRSG}$_{0.5}$ (3) (respectively in red, green, and blue). The Random Scan approach allows us to reduce the running time by a factor.}
 \label{fig:3timing2}
\end{figure}

Figure \ref{fig:3timing2} displays how the \emph{MRSG}$_{0.5}$ and \emph{AMRSG} algorithms scale with the number of nodes in comparison with the $GS$ algorithm. We can note how the Random Scan approach allows us to reduce the running time by a factor. In the case of \emph{MRSG}$_{0.5}$ and \emph{AMRSG}, the running time is, on average, reduced by 25\%.

\subsection{Dynamic Setup: Results}

\begin{figure}[!h]
  \centering
   \includegraphics[width= \textwidth]{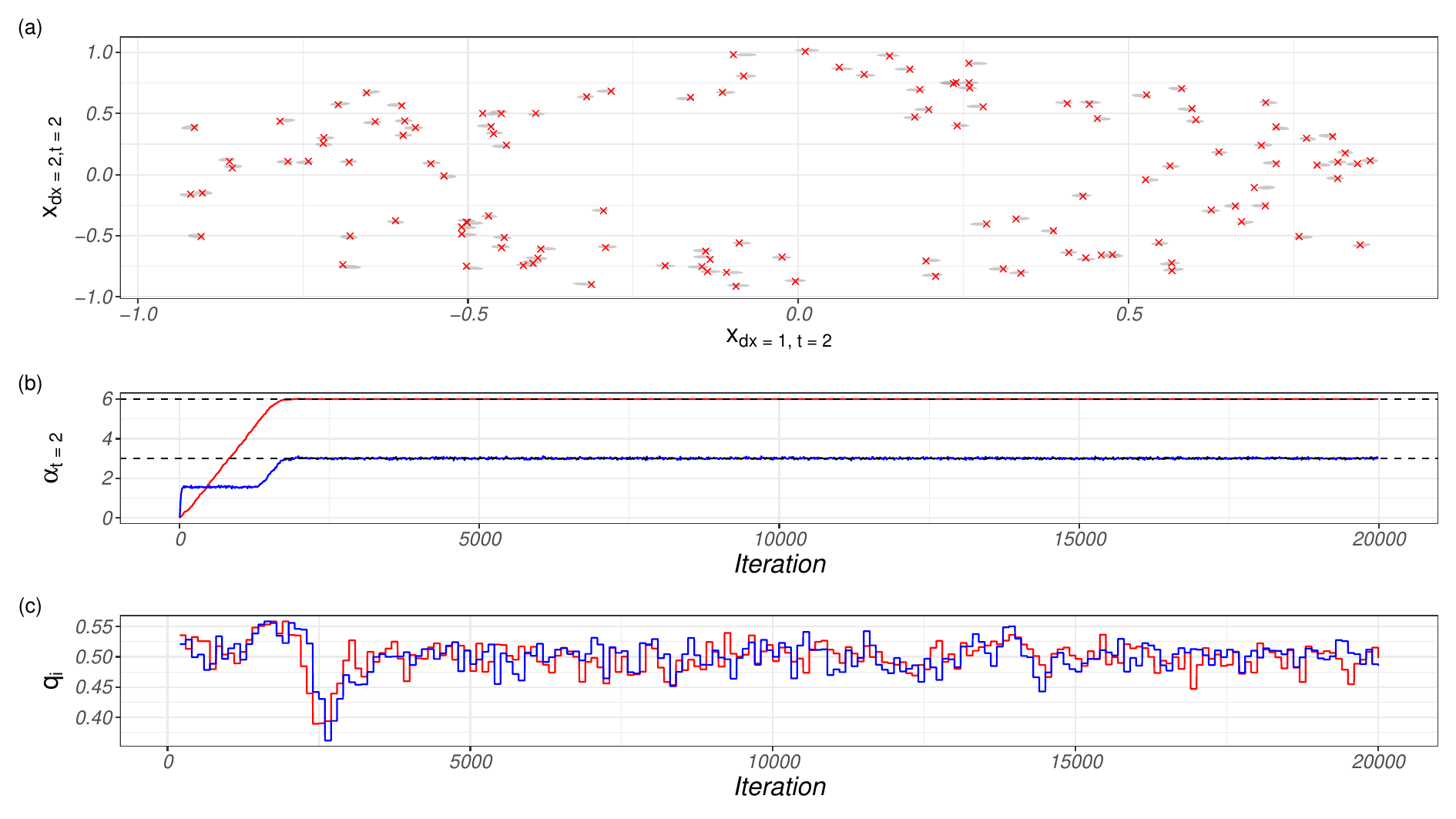}
   \caption{\textbf{Diagnostics Multi-layer Temporal layout:} Panel a) displays the estimated latent coordinates at time $t = 2$ (grey dots) against their true values (red crosses). Panel b) displays the trace plots of $\alpha_1$ and $\alpha_2$ against their true values (dashed lines). Panel c) reports an illustrative example of the adaptive selection probabilities of the first two nodes.}
 \label{fig:3diagnostics}
\end{figure}

We proceed to test the $GS$, \emph{MRSG}$_{0.5}$, and \emph{AMRSG} algorithms on the multi-layer temporal network. As a robustness check, we report in Figure \ref{fig:3diagnostics} some plots that describe the behavior of our estimates parameters against the true values using the  \emph{AMRSG} algorithm. Panel a) displays the estimated latent coordinates at time $t = 2$ (black dots) against their true values. As expected, we correctly infer the position of the nodes on the latent space. Panel b) displays the trace plots of $\alpha_1$ and $\alpha_2$ against their true values (dashed lines). Again, we correctly infer the intercept parameters of the DGP. Panel c) reports an illustrative example of the adaptive selection probabilities of the first two nodes at $t =2$. As it is expected, the adaptive selection probabilities oscillate around the value 0.5, and the oscillations get smaller and smaller as the acceptance rate gets closer to the target rate (set at $\alpha^{*}= 0.234$).

\begin{figure}[!h]
  \centering
   \includegraphics[width= \textwidth]{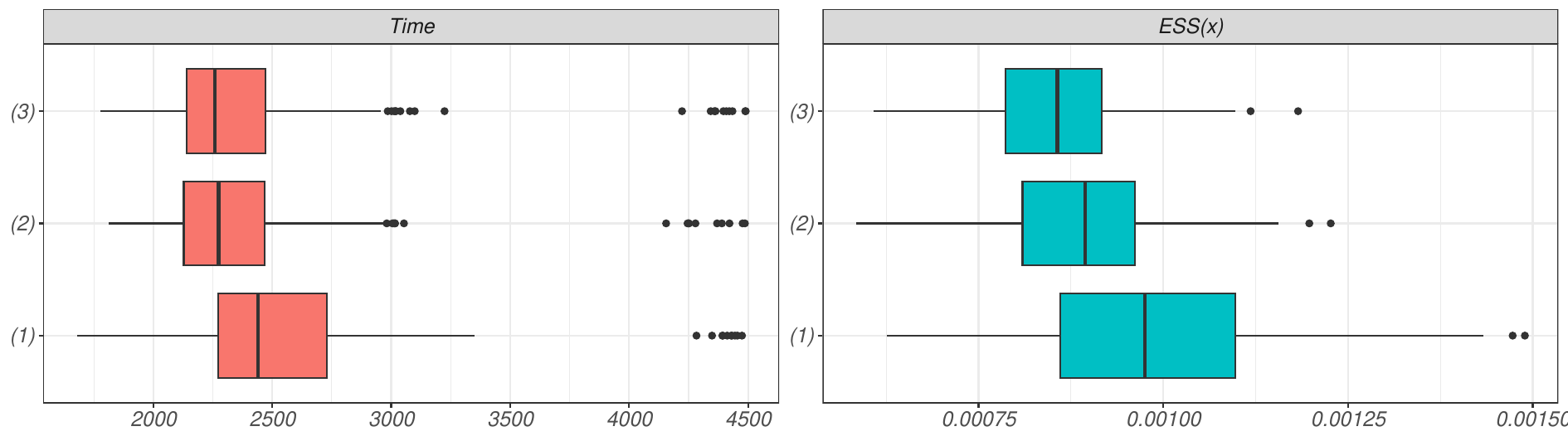}
   \caption{\textbf{Metrics - Multi-layer Temporal layout:} Comparison of $GS$ (1), \emph{AMRSG} (2) and \emph{MRSG}$_{0.5}$ (3). The boxplots report metrics comparison for 250 runs of the algorithms for the first epoch of the simulation in terms of running time (in seconds) and $\widehat{ESS}$. Although (2) and (3) have comparable running times, (2) displays a slightly better mixing (ESS).}
 \label{fig:3mt_timing}
\end{figure}

Figure \ref{fig:3mt_timing} compares $GS$, \emph{MRSG}$_{0.5}$, and \emph{AMRSG}. The boxplots report a comparison in terms of timing and $\widehat{ESS}$ for 250 runs of the algorithms and for the first epoch of the simulation. Although \emph{MRSG}$_{0.5}$ and \emph{AMRSG} have comparable running times, \emph{AMRSG} displays a slightly better mixing (higher $\widehat{ESS}$).

\section{Empirical Application}
\label{sec:emp_application}

\subsection{Data Description}
We show the effectiveness of our RS approach by applying our \emph{AMRSG} algorithm to a dataset used in a previous study by \citet{durante2017bayesian}. This dataset (see \citealp{kiti2016quantifying} for a complete description) records face-to-face interactions among people in a rural area of Kenya over three consecutive days. It includes raw contact data for 75 individuals from five households in this rural community. Each household contains multiple families living together under one head of the family. The data covers hourly face-to-face contact between individuals within each household (from 6 a.m. to 8 p.m.), but the three-day windows of data collection vary among households. Consequently, there is a lack of contact data between individuals in different households, making it impossible to analyze connectivity across households.

\begin{figure}[h!]
    \centering
            \includegraphics[width=\textwidth]{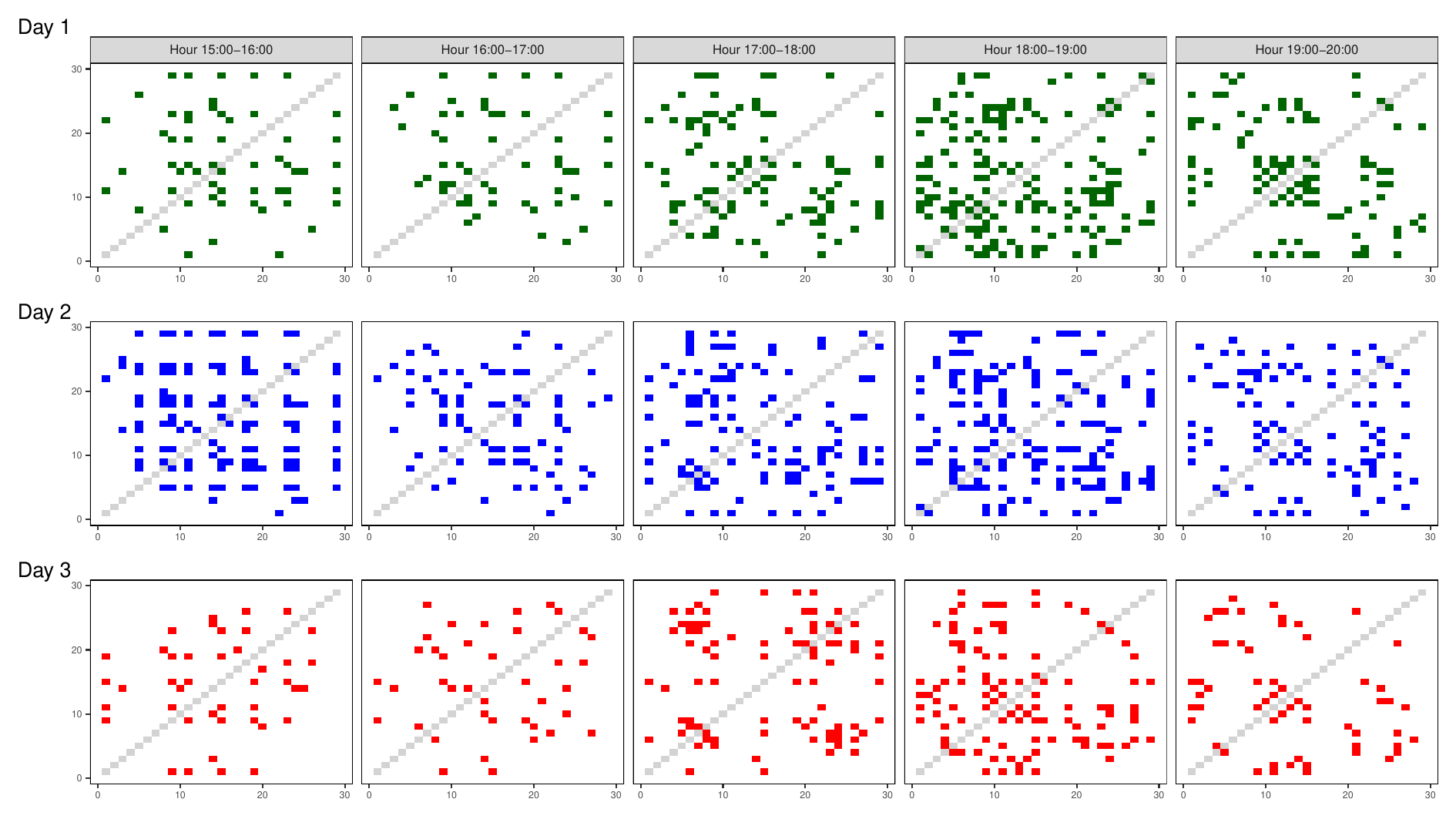}
    \caption{\textbf{Multi-layer Temporal Network} of face-to-face interactions among people in a rural area of Kenya as in \citet{kiti2016quantifying}.
    The illustrative example covers hourly face-to-face contact between individuals of the most populous household, which consists of 29 individuals, in 3 consecutive days.}
    \label{fig:3applicationMT}
\end{figure}

To avoid complications caused by the non-overlapping data collection periods, we follow the approach used by \citet{durante2017bayesian}, and we focus our analysis on the face-to-face contact networks within the most populous household, which includes 29 individuals. Figure \ref{fig:3applicationMT} reports an illustrative example of the data.

\subsection{Model}

The model we use to test our RS strategy is similar in spirit to the multi-layer temporal model presented by \citet{durante2017bayesian} although simpler. We assume $y_{ijdt} \stackrel{ind}{\sim} f_d(y_{ijdt}| g_d(\eta_{ijt}),\kappa_t)$ with day index $d=1,2,3$ and time index $t = 7 ,\ldots 20$. We assume $f_{d}$ to be the Bernoulli likelihood, $g_{d}(\eta_{ijdt}) =(1+exp(-\eta_{ijdt}))^{-1}$ the logistic link and $\eta_{ijdt} =  \alpha_{dt} - \vert\vert \mathbf{x}_{it}-\mathbf{x}_{jt}\vert\vert^2$ where $\vert\vert \mathbf{x}_{it}-\mathbf{x}_{jt}\vert\vert^2$ denotes the squared Euclidean distance. Finally, we assume the latent coordinates to follow a random-walk dynamics, i.e., $\mathbf{x}_{it} =  \mathbf{x}_{it-1} + \epsilon_{it}$, as in the simulation exercise.

\subsection{Results}

\begin{figure}[h!]
    \centering
            \includegraphics[width=\textwidth]{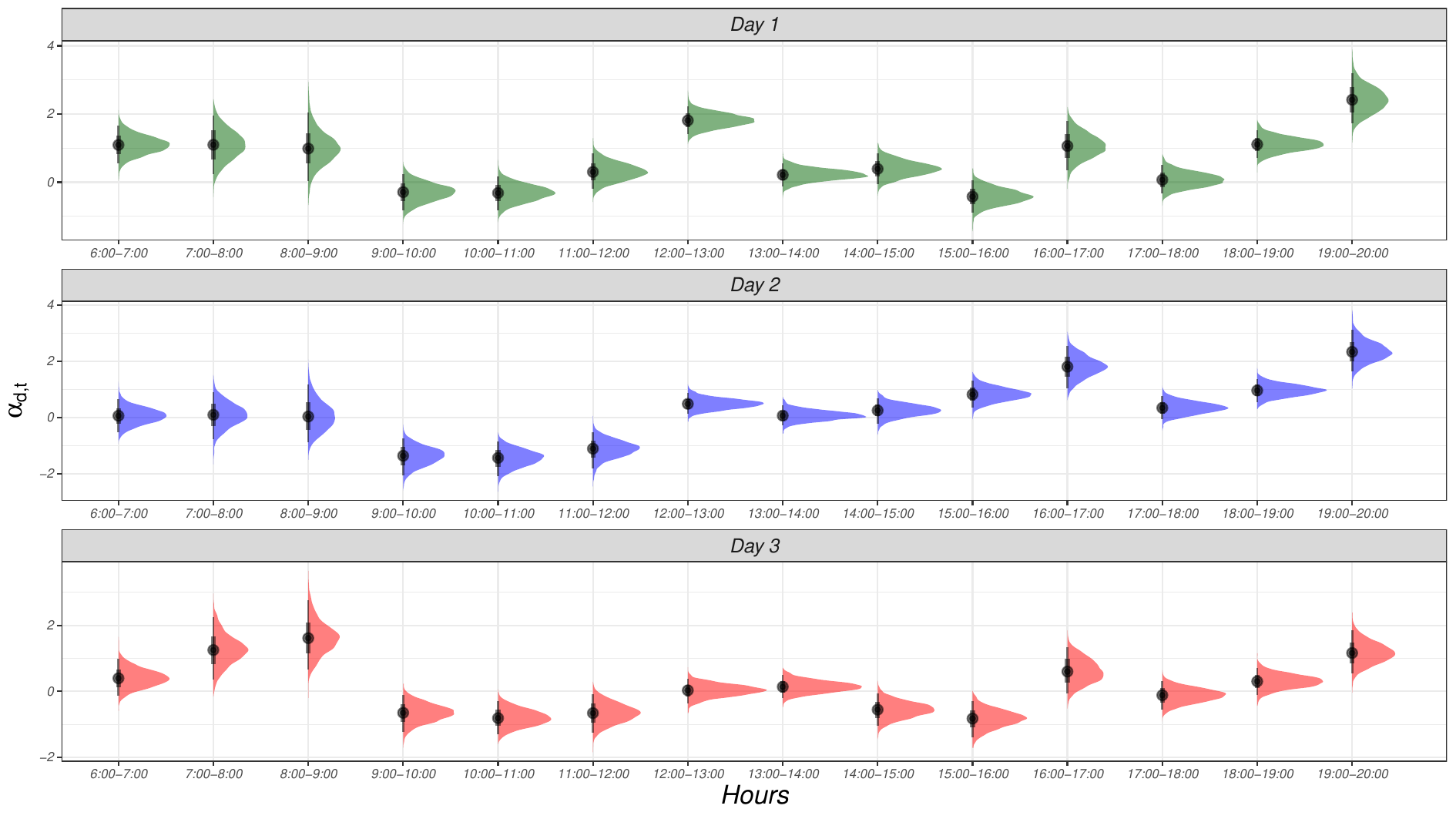}
    \caption{\textbf{Posterior of the Intercept Parameters:} The panels plots report the posterior of the parameters $\alpha_{d,t}$ obtained via the I-A algorithm for $d=1,2,3$. As expected, the intercepts get higher during meals as there are, on average, more contacts across members of the family.}
    \label{fig:3beta_time}
\end{figure}

Figure \ref{fig:3beta_time} displays the posterior of the parameters $\alpha_{d,t}$ obtained via the \emph{AMRSG} algorithm for $d=1,2,3$. The intercept parameters -- which can be interpreted as a global proxy of network concentration  (see \citealp{rastelli2016properties}) -- get higher during meal time as there are, on average, more contacts across members of the family. This result is a check of the goodness of our algorithm as it is coherent with what was found by \citet{durante2017nonparametric} although in a more complex setup.

\begin{figure}[h!]
    \centering
    \centering
            \includegraphics[width=\textwidth]{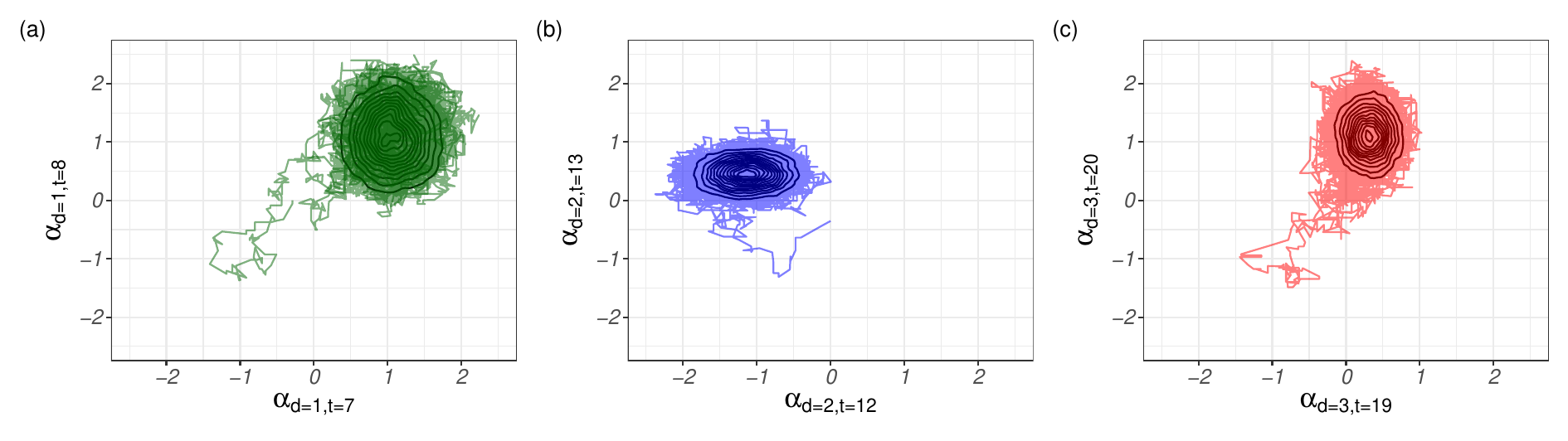}
    \caption{\textbf{Intercept Parameter Comparison:} The three plots compare the MCMC exploration of the sample space for the couples of parameters $\alpha_{d=1,t=7} \sim \alpha_{d=1,t=8}$, $\alpha_{d=2,t=12} \sim \alpha_{d=2,t=13}$ and $\alpha_{d=3,t=19} \sim \alpha_{d=2,t=20}$ (respectively in Panels a, b, and c) estimated using the \emph{AMRSG} algorithm against the same parameters estimated using AMH (black contour lines). The estimation using the random-scan procedure remains satisfactory.}
    \label{fig:3beta_comparison}
\end{figure}

We compare the estimation carried out via the $AMRSG$ algorithm against the $GS$ algorithm.
As it is known that LS models suffer from several identification issues (e.g., see \citealp{hoff2002latent}), directly comparing the estimated latent coordinates would require proper post-processing. We check whether the selected RS strategy has a severe impact on the estimation of the intercept parameter $\alpha_{d,t}$. Figure \ref{fig:3beta_comparison} compares the MCMC exploration of the sample space for the couples of parameters $\alpha_{d=1,t=7} \sim \alpha_{d=1,t=8}$, $\alpha_{d=2,t=12} \sim \alpha_{d=2,t=13}$ and $\alpha_{d=3,t=19} \sim \alpha_{d=2,t=20}$ (respectively in Panels a, b, and c) using the $AMRSG$ algorithm against the $GS$ (black contour lines). We notice that the RS algorithm 
manages to cover the same space as the $GS$ with no particular distortion. This implies that updating a subsample of latent coordinates at each iteration does not severely prejudice the estimation via the MCMC algorithm.

Finally, we compare the running time of the two algorithms. Table \ref{tab:3application} reports the running time of the two algorithms and the average $\widehat{ESS}$ computed on the intercept parameters $\alpha_{d,t}$ for each $d$ and $t$. We notice that the \emph{AMRSG} algorithm is $14.7$\% faster than the $GS$ algorithm. The price to pay for a reduction in computational time is a lower effective sample size, from 552 to 475 effective observations.

\begin{table}
\center\resizebox{0.6\textwidth}{!}{  
\begin{tabular}{cccc}
\hline\hline
\multicolumn{2}{c}{Running Time}            & \multicolumn{2}{c}{Avg $\hat{ESS}(\alpha)$}           \\ \hline
$GS$                   & \emph{AMRSG}               & $GS$                       & \emph{AMRSG}                            \\\hline

13'600.56 sec          & 11'598.75 sec        & 552 obs                & 475 obs                \\\hline
\hline
\end{tabular}}

\caption{\textbf{Application Comparison - $GS$ and \emph{AMRSG}:} The first column reports the computing time in seconds, respectively, for the $GS$ and the \emph{AMRSG} algorithms. The \emph{AMRSG} algorithm is $14.7$\% faster. The second column reports the average effective sample size (Avg $\hat{ESS}$) for the intercept parameters. The price to pay for a reduction in computational time is a lower effective sample size. The two algorithms are run for 20'000 iterations; adaptation step every 100, no burn-in, no thinning.} \label{tab:3application}
\end{table}

\section{Conclusion}
\label{sec:conclusion}
In this work, we investigated the use of the MRS approach as an MCMC strategy that alleviates the computational burden for LS models while maintaining the benefits of a general-purpose technique. We also proposed a novel adaptive random scan strategy that adjusts selection probabilities according to the acceptance rate of the MH step along with its block-adaptive version. We provided evidence that the Multiple Random-Scan approach effectively reduces the computational costs of LS models without prejudicing inference. Moreover, we showed via simulation that the Adaptive MRS approach we propose performs better than the MRS in terms of mixing. Finally, we provided an application in which we show that our Adaptive MRS implementation allows us to obtain results faster than the systematic sampler without a severe impact on parameter estimation.

\newpage
\singlespacing
\bibliographystyle{apalike}
\bibliography{biblio}
\newpage

 \bigskip
 \begin{center}
 {\large\bf SUPPLEMENTARY MATERIAL}
 \end{center}







\appendix

\renewcommand\thefigure{A.\arabic{figure}}
\setcounter{figure}{0}
\renewcommand\theequation{A.\arabic{equation}}
\setcounter{equation}{0}
\renewcommand\thetable{A.\arabic{table}}
\setcounter{table}{0}

\section{Adaptive Metropolis Hastings}
\label{apendix:amh}

\begin{algorithm}
\caption{Adaptive Metropolis-Hastings - \citealp{haario2001adaptive}}
\label{alg:haario_amh}

\begin{algorithmic}[1]
\STATE Initialize the $d$-dimensional vector $\boldsymbol{\theta}_0$
\STATE Initialize the proposal variance-covariance matrix $\Sigma_0$
\STATE Choose a probability $\beta$ small
\STATE Set the number of iterations $H$

\FOR{$h = 1$ to $H$}
    \IF{$h \leq 2d$}
    \STATE Sample $\Tilde{\boldsymbol{\theta}}_h \sim \mathcal{N}(\boldsymbol{\theta}_{h-1}, \frac{(0.1)^{2}}{d}I_{d})$
    \ELSE
     \STATE Sample \texttt{select} from $\mathcal{B}er(\beta)$
    \IF{\texttt{select} $=1$}
      \STATE Sample $\Tilde{\boldsymbol{\theta}}_h \sim \mathcal{N}(\boldsymbol{\theta}_{h-1}, \frac{(0.1)^{2}}{d}I_{d})$
    \ELSE
      \STATE Sample $\Tilde{\boldsymbol{\theta}}_h \sim \mathcal{N}(\boldsymbol{\theta}_{h-1}, \frac{2.38}{d}\boldsymbol{\Sigma}_h)$
     \ENDIF
    \ENDIF

    \STATE Compute acceptance probability: $\alpha(\boldsymbol{\theta}_{h-1}, \Tilde{\boldsymbol{\theta}}_h)$
    \STATE Generate a uniform random variable $u \sim \mathcal{U}(0,1)$
    
    \IF{$u \leq \alpha(\boldsymbol{\theta}_{h-1}, \Tilde{\boldsymbol{\theta}}_h)$}
        \STATE Set $\boldsymbol{\theta}_h = \Tilde{\boldsymbol{\theta}}_h$
    \ELSE
        \STATE Set $\Tilde{\boldsymbol{\theta}}_h = \boldsymbol{\theta}_{h-1}$
    \ENDIF
    \STATE Update $\Sigma_h$ as in \citet{haario2001adaptive}
\ENDFOR
\end{algorithmic}
\end{algorithm}

\begin{algorithm}
\caption{Adaptive Metropolis-Hastings - \citealp{roberts2009examples}}
\label{alg:roberts_amh}

\begin{algorithmic}[1]
\STATE Initialize $\boldsymbol{\theta}_0$
\STATE Set the number of iterations $H$, $\delta$, $v$ and $\alpha^{*}$

\FOR{$h = 1$ to $H$}
    \STATE Sample $\Tilde{\boldsymbol{\theta}}_h \sim \mathcal{N}(\boldsymbol{\theta}_{h-1}, e^{2\delta_{h-1}}I)$
    \STATE Compute acceptance probability: $\alpha(\boldsymbol{\theta}_{h-1}, \Tilde{\boldsymbol{\theta}}_h)$
    \STATE Generate a uniform random variable $u \sim \mathcal{U}(0,1)$
    
    \IF{$u \leq \alpha(\boldsymbol{\theta}_{h-1}, \Tilde{\boldsymbol{\theta}}_h)$}
        \STATE Set $\boldsymbol{\theta}_h = \Tilde{\boldsymbol{\theta}}_h$
    \ELSE
        \STATE Set $\Tilde{\boldsymbol{\theta}}_h = \boldsymbol{\theta}_{h-1}$
    \ENDIF

  \IF{$ [\frac{h}{v}]= 0$ }
    \IF{$\alpha(\boldsymbol{\theta}_{h-1}, \Tilde{\boldsymbol{\theta}}_h) \leq \alpha^{*}$}

     \STATE $\delta_h = \delta_{h-1} - 1/(h/v)$
     
    \ELSE

    \STATE $\delta_h = \delta_{h-1} + 1/(h/v)$

    \ENDIF
  \ENDIF

\ENDFOR

\end{algorithmic}
\end{algorithm}

\begin{algorithm}
\caption{Adaptive Metropolis-Hastings N.4 - \citealp{andrieu2008tutorial}}
\label{alg:andrieu_amh}

\begin{algorithmic}[1]
\STATE Initialize $\boldsymbol{\theta}_0$, $\boldsymbol{\mu}_0$, $\boldsymbol{\Sigma}_0$
\STATE Choose a target acceptance rate $\alpha^*$ and set $\psi \in (0,1)$
\STATE Set the number of iterations $H$
\STATE Initialize $\delta_1$, $\gamma_1 = \frac{1}{{h^\psi}}$

\FOR{$h = 1$ to $H$}
    \STATE Sample $\Tilde{\boldsymbol{\theta}}_h \sim \mathcal{N}(\boldsymbol{\theta}_{h-1}, \delta_{h}\boldsymbol{\Sigma}_{h-1})$
    \STATE Compute acceptance probability: $\alpha(\boldsymbol{\theta}_{h-1}, \Tilde{\boldsymbol{\theta}}_h)$
    \STATE Generate a uniform random variable $u \sim \mathcal{U}(0,1)$
    
    \IF{$u \leq \alpha(\boldsymbol{\theta}_{h-1}, \Tilde{\boldsymbol{\theta}}_h)$}
        \STATE Set $\boldsymbol{\theta}_h = \Tilde{\boldsymbol{\theta}}_h$
    \ELSE
        \STATE Set $\Tilde{\boldsymbol{\theta}}_h = \boldsymbol{\theta}_{h-1}$
    \ENDIF

     \STATE Update $\log(\delta_h) = \log(\delta_{h-1}) + \gamma_h[\alpha(\boldsymbol{\theta}_{h-1}, \Tilde{\boldsymbol{\theta}}_h) - \alpha^*]$
    \STATE Update $\boldsymbol{\mu}_h = \boldsymbol{\mu}_{h-1} + \gamma_h(\boldsymbol{\theta}_h - \boldsymbol{\mu}_{h-1})$
    \STATE Update $\boldsymbol{\Sigma}_h = \boldsymbol{\Sigma}_{h-1} + \gamma_h[(\boldsymbol{\theta}_h - \boldsymbol{\mu}_{h-1})(\boldsymbol{\theta}_h - \boldsymbol{\mu}_{h-1})' - \boldsymbol{\Sigma}_{h-1}]$
    
    \STATE Update $\gamma_h = \frac{1}{{h^\psi}}$
    
\ENDFOR

\end{algorithmic}
\end{algorithm}

\renewcommand\thefigure{B.\arabic{figure}}
\setcounter{figure}{0}
\renewcommand\theequation{B.\arabic{equation}}
\setcounter{equation}{0}
\renewcommand\thetable{B.\arabic{table}}
\setcounter{table}{0}

\section{Comparison with Latentnet and Stan}
\label{sec:offtheshelf}

We compare our implementation of the MCMC algorithm for LS models against the implementation adopted in the R package \texttt{latentnet} by \citet{handcock2008fitting} as well as an implementation using Stan (\citealp{carpenter2017stan}).

Although the comparison across different algorithmic implementations may be challenging (due to differences in the implemented algorithmic strategies and code optimization), we manage to show the advantages of the RS strategy for large-scale networks. We provide a comparison between our implementation of the \emph{AMRSG} algorithm and the $GS$ implementation adopted both 
 in \texttt{latentnet} and in Stan. The simulation setup consists of a random latent-coordinates layout and a standard Poisson latent-space mode with $d = 2$.

The algorithmic strategy adopted in \texttt{latentnet} consists of an adaptation in the Metropolis step for the burn-in phase and a regular MH afterward with a proposal centered around previously obtained posterior modes. We modify our \emph{AMRSG} algorithm to accommodate such a feature. Moreover, we choose $c = 1$ and choose $u = 5000$ to update the selection probabilities every 5'000 iterations. We run both algorithms 10 times for each setup consisting of different network sizes $N \in \{250, 500, 1000, 1250, 1500\}$ with 50'000 iterations and use the first 10'000 as a burn-in. We track computing times for the two algorithms, the mean squared error, and the variance of the chains. The algorithmic strategy adopted in Stan is a No-U-turn Hamiltonian Monte Carlo. We run the algorithm 5 times for each setup with 5'000 iterations for $N \in \{250, 500, 1000\}$. 

\begin{figure}[!htb]
    \centering
    \includegraphics[width= \textwidth]{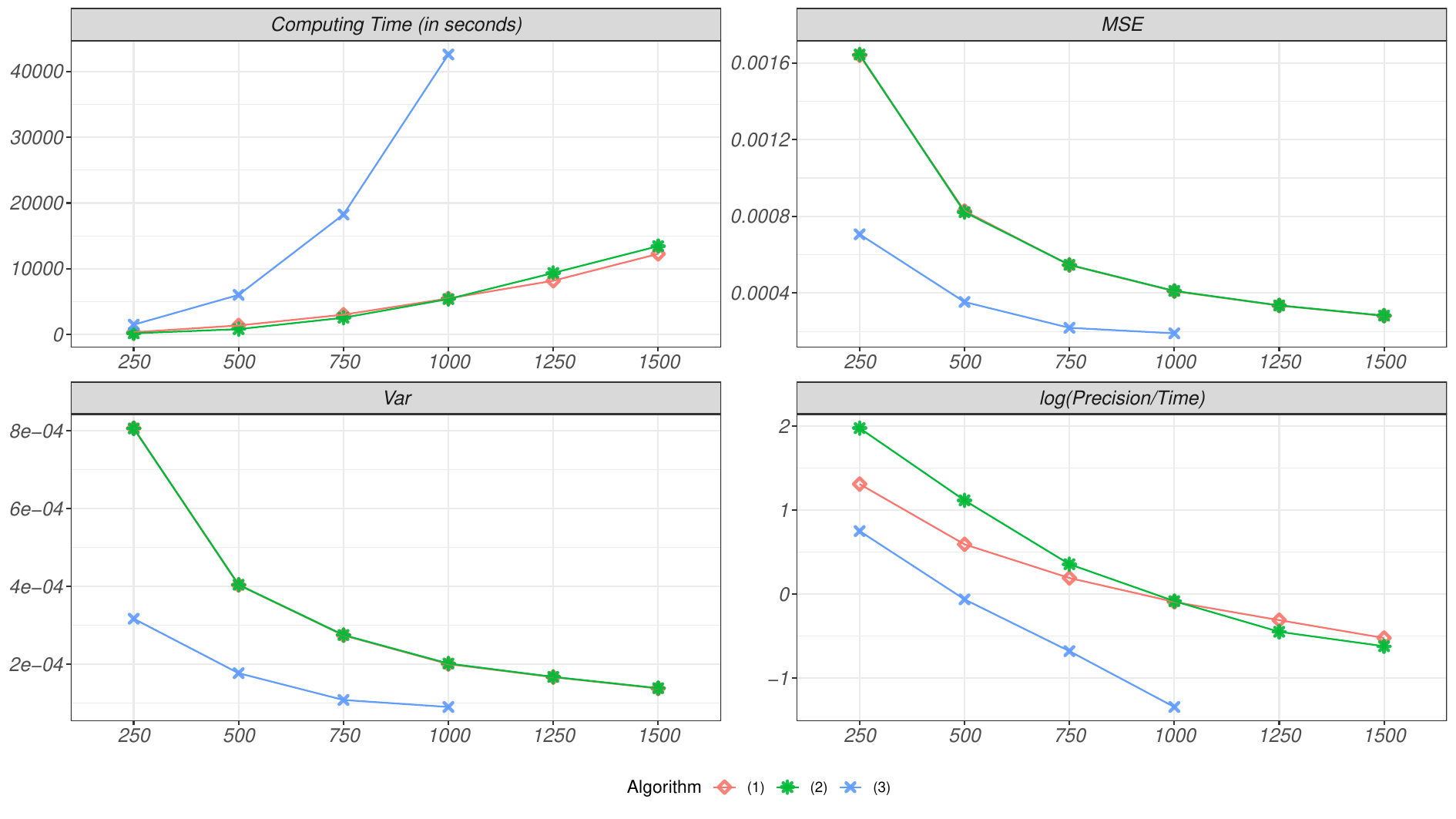}
    \caption{ \textbf{\ Latentnet and Stan comparison:} The performance of our \emph{AMRSG} (1) in red, the R package \texttt{latentnet} (2) in green, and the Stan algorithmic implementation in blue (3). The panels report the Computing Time in seconds (top-left), MSE (top-right), Variance (bottom-left), and log(Precision/Time) (bottom-right) for an increasing number of nodes (horizontal axis). Each dot is an average across 10 algorithmic iterations for each $\emph{AMRSG}$ and \texttt{latentnet} algorithms, while across 5 iterations for Stan, due to RAM budget.}
    \label{fig:timing_latentnet_stan}
\end{figure}

Figure \ref{fig:timing_latentnet_stan} displays the comparison between the three algorithms in terms of computational time (top-left panel), MSE (top-right panel), Variance (bottom-left panel), and precision-to-time ratio in log scale (bottom right), where we defined precision as the inverse of the Variance. Each dot is an average across the algorithmic iterations for each of the three algorithms.

Regarding computational time, the Stan algorithm is the slowest, while the algorithm implemented in \texttt{latentnet} is well-optimized, as it turns out to be faster for medium-sized networks (below 1'000 nodes). On the other hand, our \emph{AMRSG} algorithm shows its advantages for large-sized networks (above 1'000 nodes). Boxplot inspection reveals that the difference in computational time is significant between \emph{AMRSG} and \texttt{latentnet}. \texttt{latentnet} and   \emph{AMRSG} are comparable in terms of MSE and Variance, while the Stan algorithm exhibits a better performance. Finally, the precision-to-time ratio is better for \texttt{latentnet} for medium-sized networks, but \emph{AMRSG} proves to be better as the network nodes increase. We want to stress that these MCMC strategies are complements rather than substitutes. This means that one could combine the speed of a highly optimized algorithm with the improvement in scalability of an RS strategy.

\begin{figure}[!htb]
    \centering
    \includegraphics[width= \textwidth]{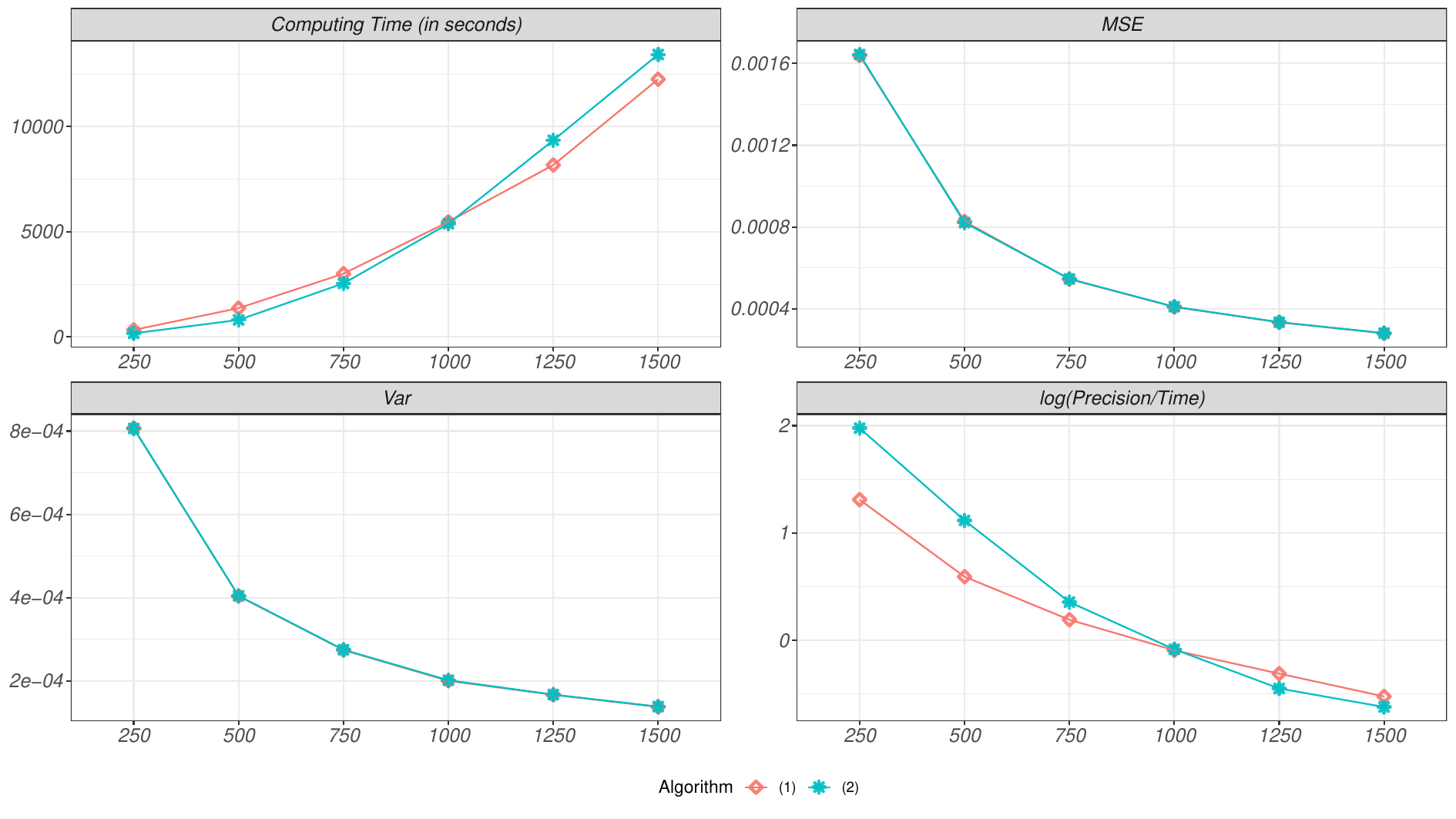}
    \caption{ \textbf{\ Latentnet comparison:} The performance of our \emph{AMRSG} (1) in red and the R package \texttt{latentnet} (2) in green. The panels report the Computing Time in seconds (top-left), MSE (top-right), Variance (bottom-left), and log(Precision/Time) (bottom-right) for an increasing number of nodes (horizontal axis). Each dot is an average across 10 algorithmic iterations for each of the two algorithms.}
    \label{fig:timing_latentnet}
\end{figure}

\clearpage

\renewcommand\thefigure{C.\arabic{figure}}
\setcounter{figure}{0}
\renewcommand\theequation{C.\arabic{equation}}
\setcounter{equation}{0}
\renewcommand\thetable{C.\arabic{table}}
\setcounter{table}{0}

\section{More Results}
\label{apx:rand}

\begin{figure}[!htb]
  \centering
   \includegraphics[width= \textwidth]{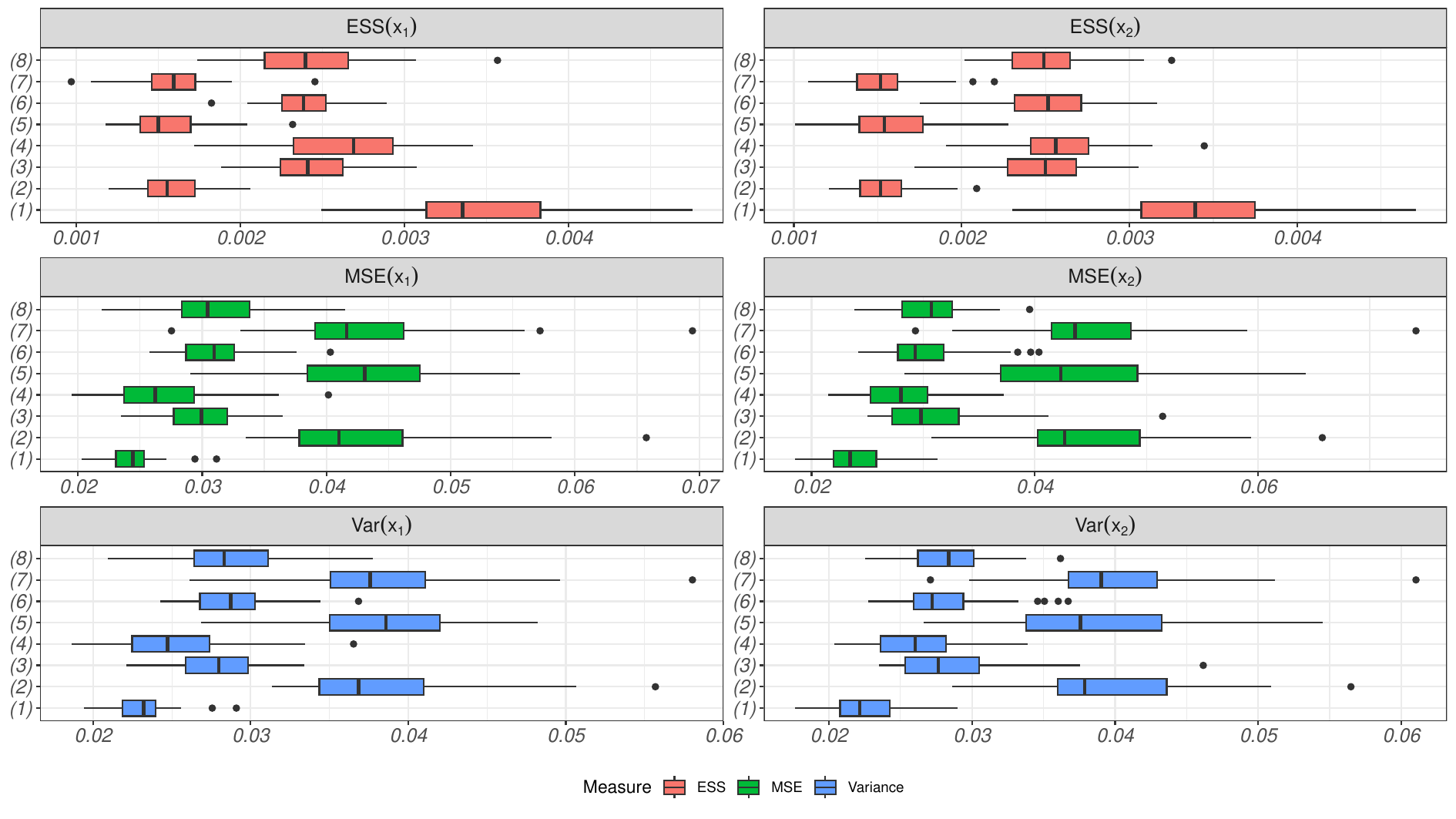}
   \caption{\textbf{Metrics - No Burn-in and Thinning with 5'000 iterations:} Comparison between the competing algorithms. The boxplots report metrics comparison for 50 runs of the algorithms in \ref{subsec:algorithms}. The reported metrics are the Effective Sample Size (ESS) as a proportion of the overall sample, the Mean Squared Error (MSE) compared to the true value of the latent coordinates, and the Variance of the chains. The metrics are averaged across nodes for each latent coordinate $\mathbf{x}_1$ and $\mathbf{x}_2$. The number of iterations has been set to 5'000. The adaptive selection probabilities get updated every 100 iterations. Legend: (1) GS, (2) \emph{MRSG}$_{0.25}$, (3) \emph{MRSG}$_{0.5}$, (4) \emph{AMRSG}, (5) \emph{B-MRSG}$_4$, (6) \emph{B-MRSG}$_2$, (7)  \emph{B-AMRSG}$_4$, and (8) \emph{B-AMRSG}$_2$.}
 \label{fig:metrics_few_nodes}
\end{figure}

\begin{figure}[!htb]
  \centering
   \includegraphics[width= \textwidth]{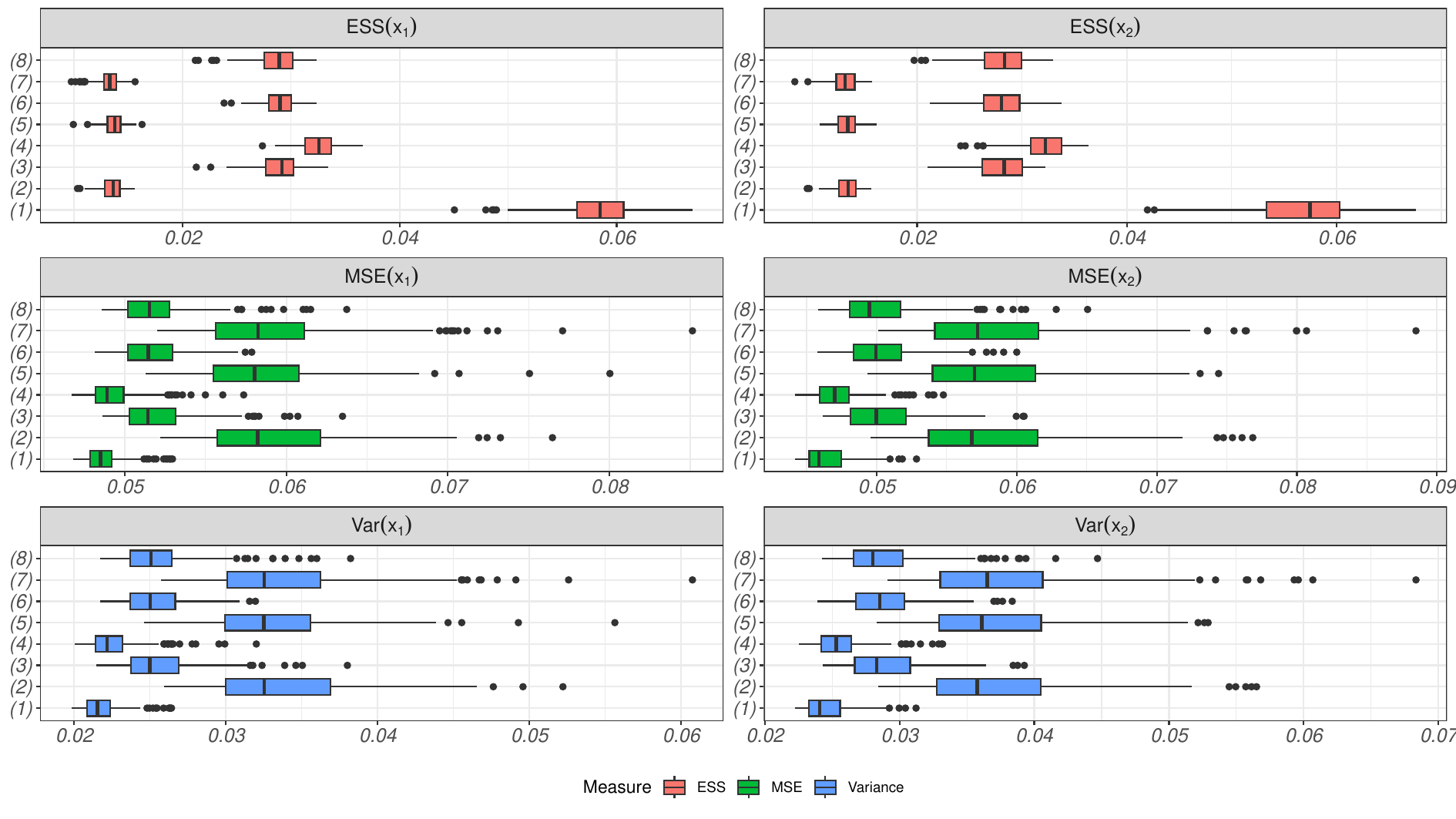}
   \caption{\textbf{Metrics - No Burn-in and Thinning on Random-Layout Binary Network:} Comparison between the competing algorithms. The boxplots report metrics comparison for 250 runs of the algorithms in \ref{subsec:algorithms}. The reported metrics are the Effective Sample Size (ESS) as a proportion of the overall sample, the Mean Squared Error (MSE) compared to the true value of the latent coordinates, and the Variance of the chains. The metrics are averaged across nodes for each latent coordinate $\mathbf{x}_1$ and $\mathbf{x}_2$. The number of iterations has been set to 30'000. The adaptive selection probabilities get updated every 100 iterations. Legend: (1) GS, (2) \emph{MRSG}$_{0.25}$, (3) \emph{MRSG}$_{0.5}$, (4) \emph{AMRSG}, (5) \emph{B-MRSG}$_4$, (6) \emph{B-MRSG}$_2$, (7)  \emph{B-AMRSG}$_4$, and (8) \emph{B-AMRSG}$_2$.}
 \label{fig:metrics_noburnin_random}
\end{figure}

\begin{figure}[!htb]
  \centering
   \includegraphics[width= 0.8\textwidth]{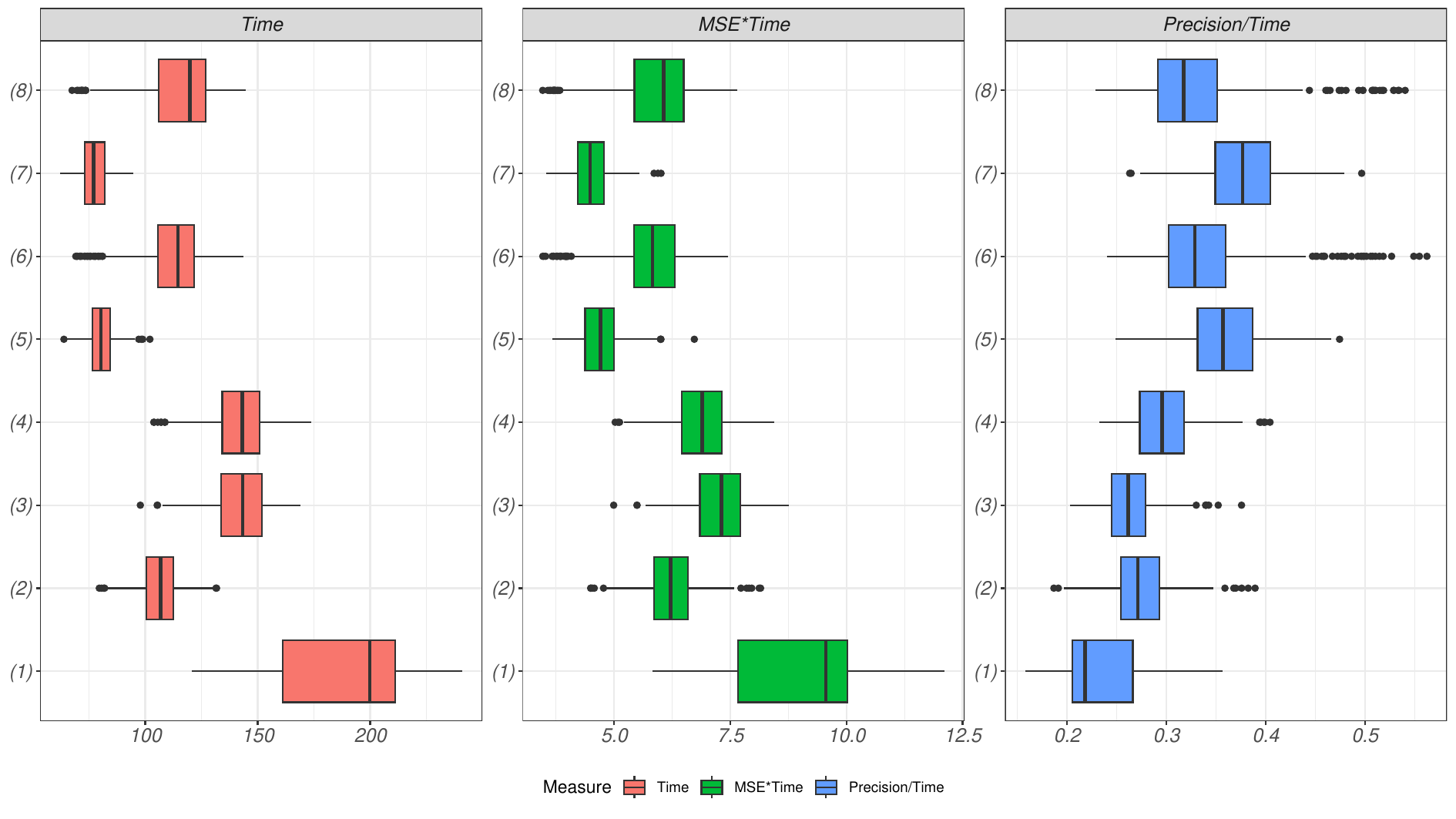}
   \caption{\textbf{Timing Random-Layout Binary Network: } Comparison between the competing models in terms of computing time in seconds, MSE, and $Time*MSE$. The boxplots report metrics comparison for 250 runs of the algorithms in \ref{subsec:algorithms}. The metrics are averaged across nodes and coordinates. The adaptive selection probabilities get updated every 100 iterations. Legend: (1) GS, (2) \emph{MRSG}$_{0.25}$, (3) \emph{MRSG}$_{0.5}$, (4) \emph{AMRSG}, (5) \emph{B-MRSG}$_4$, (6) \emph{B-MRSG}$_2$, (7)  \emph{B-AMRSG}$_4$, and (8) \emph{B-AMRSG}$_2$.}
 \label{fig:timing_rand}
\end{figure}

\begin{figure}[!htb]
  \centering
   \includegraphics[width= \textwidth]{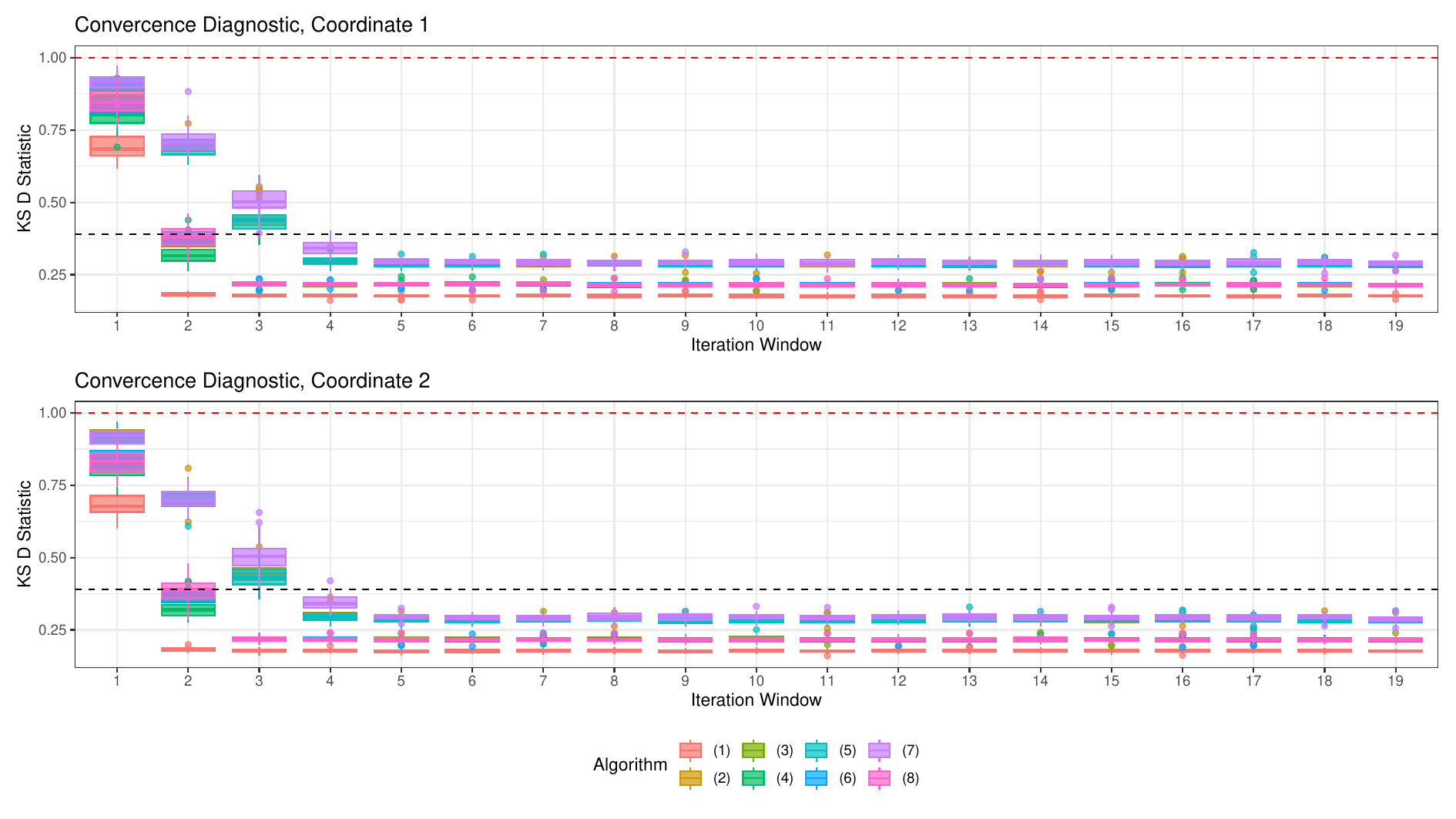}
   \caption{\textbf{Kolmogorov-Smirnov Diagnostic.} The two charts report the boxplots of the Kolmogorov-Smirnov $D$ statistic obtained by performing the test between a sequence of non-overlapping 500-iteration chain subsamples and the last obtained chain subsample with thinning every 10.  The D statistic has been averaged across nodes and is presented separately for the first and second coordinates (top and bottom panels). The algorithms have been run for 10'000 iterations without burn-in. The charts report the maximum value attainable by the $D$ (red dashed line) and the critical value of the statistic $D^{*}_{\alpha, n}$ with significance level $\alpha = 0.01$ (black dashed line). Legend: (1) $GS$, (2)  \emph{B-AMRSG}$_2$, (3) \emph{B-AMRSG}$_4$, (4) \emph{B-MRSG}$_2$, (5) \emph{B-MRSG}$_4$, (6) \emph{AMRSG}, (7) \emph{MRSG}$_{0.25}$, (8) \emph{MRSG}$_{0.5}$.}
 \label{fig:ks_diagnostic}
\end{figure}

\end{document}